\documentclass[aps,pra,twocolumn,superscriptaddress]{revtex4-1}

\usepackage{graphicx}
\usepackage{dcolumn}
\usepackage{bm}
\usepackage{amsmath}
\usepackage{amssymb}
\usepackage{latexsym}
\usepackage{epsfig}
\usepackage{amsbsy}
\usepackage{array}
\usepackage{amssymb}
\usepackage{setspace}
\usepackage{bm}
\usepackage{epstopdf}
\usepackage{subfigure}
\usepackage{xcolor}

\def\sint{\ifmmode{- \!\!\!\!\!\! \int}
    \else{\hbox{$- \!\!\!\! \int \ $}}\fi}

\begin{document}

\title{Large amplitude mechanical coherent states and detection of weak nonlinearities in cavity optomechanics}


\author{Wenlin Li}
\thanks{These two authors contributed equally}
\affiliation{School of Science and Technology, Physics Division, University of Camerino, I-62032 Camerino (MC), Italy}
\affiliation{College of Sciences, Northeastern University, Shenyang 110819, China}
\author{Paolo~Piergentili}
\thanks{These two authors contributed equally}
\affiliation{School of Science and Technology, Physics Division, University of Camerino, I-62032 Camerino (MC), Italy}
\affiliation{INFN, Sezione di Perugia, I-06123 Perugia, Italy}
\author{Francesco Marzioni}
\affiliation{School of Science and Technology, Physics Division, University of Camerino, I-62032 Camerino (MC), Italy}
\affiliation{INFN, Sezione di Perugia, I-06123 Perugia, Italy}
\author{Michele Bonaldi}
\affiliation{Institute of Materials for Electronics and Magnetism, Nanoscience-Trento-FBK Division, 38123 Povo, Trento, Italy}
\affiliation{Istituto Nazionale di Fisica Nucleare (INFN), Trento Institute for Fundamental Physics and Application, I-38123 Povo,
Trento, Italy}
\author{Antonio Borrielli}
\affiliation{Institute of Materials for Electronics and Magnetism, Nanoscience-Trento-FBK Division, 38123 Povo, Trento, Italy}
\author{Enrico Serra}
\affiliation{Institute of Materials for Electronics and Magnetism, Nanoscience-Trento-FBK Division, 38123 Povo, Trento, Italy}
\affiliation{Istituto Nazionale di Fisica Nucleare (INFN), Trento Institute for Fundamental Physics and Application, I-38123 Povo,
Trento, Italy}
\author{Francesco Marin}
\affiliation{INFN, Sezione di Firenze, I-50125 Firenze, Italy}
\affiliation{CNR-INO, L.go Enrico Fermi 6, I-50125 Firenze, Italy}
\affiliation{Dipartimento di Fisica e Astronomia, Universita di Firenze, Via Sansone 1, I-50019 Sesto Fiorentino (FI), Italy}
\affiliation{European Laboratory for Non-Linear Spectroscopy (LENS), Via Carrara 1, I-50019 Sesto Fiorentino (FI), Italy}
\author{Francesco Marino}
\affiliation{CNR-INO, L.go Enrico Fermi 6, I-50125 Firenze, Italy}
\author{Nicola~Malossi}
\affiliation{School of Science and Technology, Physics Division, University of Camerino, I-62032 Camerino (MC), Italy}
\affiliation{INFN, Sezione di Perugia, I-06123 Perugia, Italy}
\author{Riccardo~Natali}
\affiliation{School of Science and Technology, Physics Division, University of Camerino, I-62032 Camerino (MC), Italy}
\affiliation{INFN, Sezione di Perugia, I-06123 Perugia, Italy}
\author{Giovanni~Di~Giuseppe}
\affiliation{School of Science and Technology, Physics Division, University of Camerino, I-62032 Camerino (MC), Italy}
\affiliation{INFN, Sezione di Perugia, I-06123 Perugia, Italy}
\author{David~Vitali}
\affiliation{School of Science and Technology, Physics Division, University of Camerino, I-62032 Camerino (MC), Italy}
\affiliation{INFN, Sezione di Perugia, I-06123 Perugia, Italy}
\affiliation{CNR-INO, L.go Enrico Fermi 6, I-50125 Firenze, Italy}

\date{\today}
\begin{abstract}
The generation of large-amplitude coherent states of a massive mechanical resonator, and their quantum-limited detection represent useful tools for quantum sensing and for testing fundamental physics theories. In fact, any weak perturbation may affect the coherent quantum evolution of the prepared state, providing a sensitive probe for such a perturbation. Here we consider a cavity optomechanical setup and the case of the detection of a weak mechanical nonlinearity. We consider different strategies, first focusing on the stationary dynamics in the presence of multiple tones driving the system, and then focusing on non-equilibrium dynamical strategies. These methods can be successfully applied for measuring Duffing-like material nonlinearities, or effective nonlinear corrections associated with quantum gravity theories. 
\end{abstract}
\maketitle

\section{Introduction}

Cavity optomechanics, where one or more microwave or optical cavity modes interact dispersively with one or more mechanical resonators~\cite{Aspelmeyer2014}, has allowed the generation of various examples of nonclassical states of macroscopic mechanical resonators, namely squeezed states~\cite{Wollman2015,Pirkkalainen2015,Lecocq2015,Youssefi2023,Rossi2024}, entangled single-phonon states~\cite{Riedinger}, Gaussian bipartite entangled states~\cite{Sillanpaa,Kotler}, and Schr\"odinger cat states~\cite{Chu}. Coherent states of a massive mechanical oscillator seem instead less interesting because they do not show any non-classical feature, such as sub-shot noise, negative quasiprobabilities, or quantum correlations. However, the simplicity of their quantum dynamics allows one to detect any weak perturbation of the resonator Hamiltonian able to modify the amplitude or phase of the free coherent dynamics. This is particularly useful when testing theories concerning the unknown territory between quantum mechanics and gravity, such as those associated with deformed commutators in the nonrelativistic limit~\cite{Pikovski,Bawaj2015,bushev,bonaldi,kumar,tobar,Bosso}, those aiming at verifying the quantum nature of the gravitational field \cite{bose,marletto,lami,bosermp,marlettormp}, or nonlocal approaches to quantum gravity~\cite{belenchia,Wang}. In fact, new physics may manifest itself through a modification of the harmonic evolution of the mechanical resonator, acting as an effective dynamical nonlinearity. Therefore, monitoring the time evolution of an initially prepared coherent state represents a powerful way of detecting this effective tiny nonlinearity, limited only by the quantum zero-point fluctuations. In particular, large-amplitude coherent states become extremely sensitive probes whenever the dynamical perturbation affects the phase dynamics of the resonator.

Here we analyze how to adjust ground state cooling in order to prepare the mechanical resonator in a large amplitude coherent state, and then discuss various strategies for measuring an effective Duffing nonlinearity, which can be either of material/geometric origin~\cite{Lifshitz}, or the effective nonlinear corrections associated with quantum gravity theories, such as those yielding deformed commutation rules in the non-relativistic limit~\cite{Pikovski,Bawaj2015,bushev,bonaldi,kumar,tobar,Bosso}, and quantified by an effective nonlinearity parameter $\beta_{NL}$. The corresponding estimation of the quantum gravity deformation parameter would be performed in a quantum regime dominated only by the zero-point fluctuations of the mechanical resonator~\cite{bonaldi}. In fact, the estimations of $\beta_{NL}$ carried out up to now~\cite{Bawaj2015,bushev,kumar,tobar}, based on mechanical resonators of different size and kind, have been mostly made in a classical regime dominated by thermal fluctuations. A notable exception is the recent study of Ref.~\cite{Donadi25}, which exploits a superconducting qubit to prepare, control, and readout a 16 $\mu$g mechanical resonator—a vibrational mode localized within a bulk sapphire crystal—in nonclassical states of motion (energy eigenstates and their superpositions), and investigated their time evolution.

The paper is organized as follows. In Sec. II we describe the cavity optomechanics setup and its dynamics in terms of Heisenberg-Langevin equations obtained from the system Hamiltonian and including dissipative and noise terms for both the mechanical and optical degrees of freedom. In Sec. III we provide an effective dynamical description in terms of two coupled set of equations: semiclassical nonlinear equations for the mechanical and optical amplitudes, and a set of linearized Quantum Langevin equations for the quantum fluctuations around the semiclassical evolution. In Sec. IV we describe the stationary state of the system determined by the driving fields, and discuss the possibility to estimate the small nonlinear coefficient from such a steady state. In Sec. V we discuss a more efficient and unbiased scheme for estimating the nonlinearity, based on monitoring the nonstationary decay of the mechanical resonator to the thermal equilibrium, starting from the stationary state described in Sec. IV. Sec. VI is for concluding remarks, while in the Appendix we discuss an alternative nonstationary strategy, sligthly different form that of Sec. V.

\section{The general model}  

\begin{figure}[!b]
	\centering	\includegraphics[width=.99\linewidth]{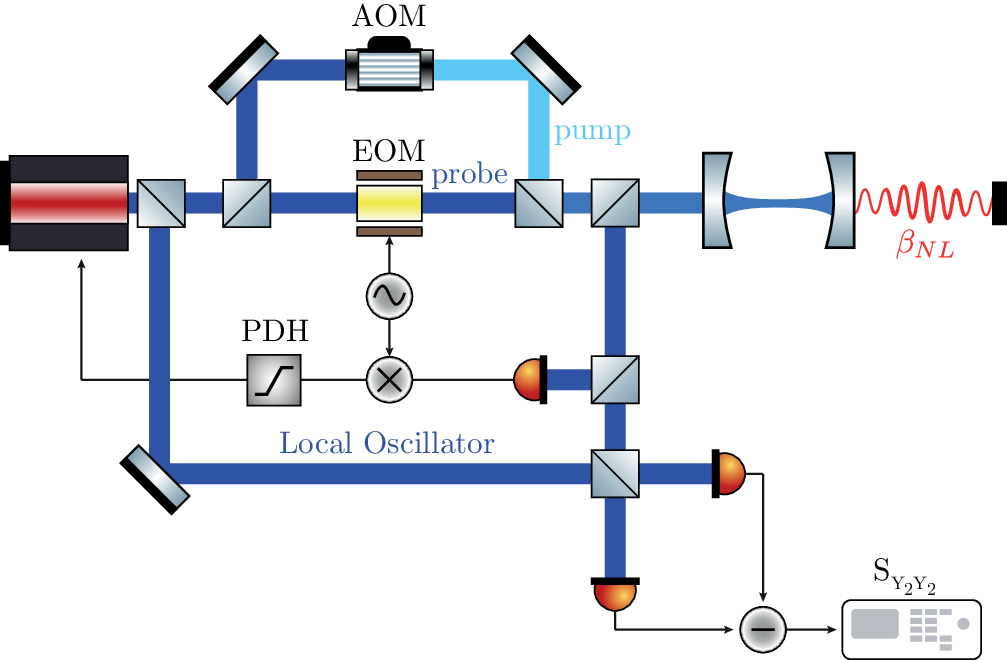}
	\caption{
Scheme of the optomechanical setup. A probe beam, with carrier frequency $\omega_{L2}$ impinges an optomechanical cavity with a movable end mirror. The EOM frequency modulates the probe beam to lock the probe in resonance with the optomechanical cavity using the Pound-Drever-Hall (PDH) technique. The reflected probe beam is analyzed by homodyne detection to measure the mechanical motion of the movable mirror and estimate the nonlinear parameter $\beta_{NL}$, which is highlighted in the picture by a deformed spring attached to the movable mirror. A second beam, pump beam, with a carrier frequency $\omega_{L1}$ and modulation $\delta_m$, also impinges the optomechanical cavity to manipulate and drive the mechanical oscillator.  
	}
\label{Sketch}
\end{figure}

The simplest cavity optomechanical system is formed by a driven optical or microwave cavity coupled by a radiation pressure-like interaction with a mechanical resonator. A plethora of variations of this paradigmatic system have been explored in the literature, considering multimode systems, multi-tone driving, and also the eventual inclusion of nonlinear effects~\cite{Aspelmeyer2014,Genes2009,Piergentili2022}. Here we describe a general treatment of the quantum dynamics of such systems in the presence of two additional elements: a mechanical nonlinearity, and the presence of an optical (or microwave) probe mode which is used to provide a continuous, real-time detection of the mechanical motion. A schematic description of the system we are going to study is provided by Fig.~\ref{Sketch}.

The system is described by the following Hamiltonian, decomposed as follows
\begin{equation}\label{ham0}
  H=H_{\rm pump}+H_{\rm probe}+H_{\rm mech}+H_{\rm int},
\end{equation}
where
\begin{eqnarray}
  H_{\rm pump} &=& \hbar \omega_{c1}a_1^\dagger a_1+i \hbar E_1\left(a_1^\dagger e^{-i\omega_{L1}t}- a_1 e^{i\omega_{L1}t}\right)  \\
  && +i \hbar E_m\left(a_1^\dagger e^{-i(\omega_{L1}+\delta_m)t}- a_1 e^{i(\omega_{L1}+\delta_m)t}\right), \\
  H_{\rm probe} &=& \hbar \omega_{c2}a_2^\dagger a_2+i \hbar E_2\left(a_2^\dagger e^{-i\omega_{L2}t}- a_2 e^{i\omega_{L2}t}\right)  \\
  H_{\rm mech} &=& \hbar \omega_{m}b^\dagger b  + \hbar \omega_{m}\frac{\beta_{\rm NL}}{12}(be^{i\varphi}-b^\dagger e^{-i\varphi})^4, \\
  H_{\rm int} &=& -\sum_{j=1,2}\hbar g_j a_j^\dagger a_j(b+b^\dagger).
\end{eqnarray}
One has a pump cavity mode with bosonic annihilation operator $a_1$, which is bichromatically driven, i.e., it has a carrier at frequency $\omega_{L1}$ and a second tone generated by a modulation at frequency $\delta_m$ which will be used to manipulate and drive the mechanical resonator through its beat notes. For example, if the modulation frequency $\delta_m$ is quasi-resonant with the mechanical frequency $\omega_m$ the mechanical resonator is excited to a state with a nonzero coherent amplitude, which in a fully quantum regime would approach a coherent state. Alternatively, if $ \delta_m \sim 2 \omega_m$, one has a parametric modulation which, again in the quantum regime, is able to generate squeezing of the mechanical resonator \cite{Chowdury2020}.

The probe mode described by the bosonic annihilation operator $a_2$, is driven, in general, at a different frequency $\omega_{L2}$, and it refers to a different cavity mode from the one driven by the pump (different frequency and/or polarization) in order to avoid interference between the two drivings. The driving rates are explicitly given by $E_j=\sqrt{2\kappa_{j,in} P_j/\hbar\omega_{Lj}}$, $j=1,2$, with $\kappa_{j,in}$ the j-th cavity mode decay rate through the input port, and $P_j$ the associated laser input power.

The mechanical mode Hamiltonian is described by means of the annihilation mechanical operator $b$, and is characterized by a fourth-order nonlinearity, quantified in terms of a dimensionless nonlinearity parameter $\beta_{\rm NL}$. This nonlinear term can be associated with a mechanical Duffing nonlinearity, quartic in the position variable ($\varphi =\pi/2$), which is the lowest-order deviation from the harmonic potential in parity invariant materials~\cite{Lifshitz}. Alternatively, it can describe the effective nonlinearity associated with deformed commutator phenomenological theories of quantum gravity, which is quartic in momentum ($\varphi =0$, see Refs.~\cite{Pikovski,Bawaj2015,bushev,bonaldi,kumar,tobar,Bosso,bose} and references therein). Finally we have the usual radiation pressure dispersive interaction between the pump and probe modes with the mechanical mode, quantified by the single-photon optomechanical coupling rates $g_j=-(d\omega_{cj}/dx)x_{\rm zpf}$, where $x_{\rm zpf}=\sqrt{\hbar/2m\omega_m}$ is the spatial width of the oscillator zero-point motion, and $m$ is the resonator mass.

We then move to the interaction picture with respect to the optical Hamiltonian $H_0= \hbar \omega_{L1}a_1^\dagger a_1+\hbar \omega_{L2}a_2^\dagger a_2$, which means considering the frame rotating at the laser driving frequency for both pump and probe modes. The mechanical resonator and the cavity modes are coupled to their corresponding thermal reservoir at temperature $T$ through fluctuation-dissipation processes, which we include in the Heisenberg picture by adding dissipative and noise terms, yielding the following quantum Langevin equations~\cite{Giovannetti2001,Aspelmeyer2014}
\begin{subequations}\label{eq:quantum langevin}
\begin{align}
\dot{a_j}=&(-\kappa_j+i\Delta^{(0)}_j)a_j+E_j+\delta_{j1}E_m e^{-i\delta_{m}t}\nonumber \\
&+ig_j(b+b^\dagger)a_j+\sqrt{2\kappa_{j,in}}a_{j,in}+\sqrt{2\kappa_{j,ex}}a_{j,ex}, \label{eq:quantum langevin1}\\
\dot{b}=&(-\gamma_m-i\omega_m) b+i e^{-i\varphi}\omega_{m}\frac{\beta_{\rm NL}}{3}(b e^{i\varphi}-b^\dagger e^{-i\varphi})^3 \nonumber \\
& +i\sum_{j=1,2}   g_ja_j^\dagger a_j+\sqrt{2\gamma_m}b_{in},
\label{eq:quantum langevin2}
\end{align}
\end{subequations}
where $\delta_{j1}$ is the Kronecker delta, $\Delta^{(0)}_j=\omega_{Lj}-\omega_{cj}$, $\kappa_j=\kappa_{j,in}+\kappa_{j,ex}$ is the total cavity amplitude decay rate, $\kappa_{j,ex}$ is the optical loss rate through all the ports different from the input one, and $\gamma_m$ is the mechanical amplitude decay rate. $a_{j,in}(t)$, $a_{j,ex}(t)$  and $b_{in}$ are the corresponding zero-mean noise reservoir operators, which are all uncorrelated from each other and can be assumed to be Gaussian and white. In fact, they possess the correlation functions $\langle f(t)^\dagger f(t')\rangle = \bar{n}_f\delta(t-t')$ and $\langle f(t) f(t')^\dagger\rangle=(\bar{n}_f+1)\delta(t-t')$ where $f(t)$ is either $a_{j,in}(t)$, $a_{j,ex}(t)$ or $b_{in}$, and $\bar{n}_f=[\exp(\hbar\omega_f/k_bT)-1]^{-1}$ is the mean thermal excitation number for the corresponding mode.

\section{Semiclassical and quantum fluctuation dynamics}

Eqs.~(\ref{eq:quantum langevin}) provide the exact description of the quantum dynamics of the system. They are hard to solve because of their nonlinear nature stemming from the radiation pressure and the mechanical nonlinearities. This can be seen for example by looking at the dynamics of the expectation values of the system, which is obtained by averaging over thermal and quantum noises, i.e., by tracing Eqs.~(\ref{eq:quantum langevin}) over system and reservoir variables,
\begin{subequations}\label{eq:semiclassical}
\begin{align}
\dot{\alpha_j}=&(-\kappa_j+i\Delta^{(0)}_j)\alpha_j+ig_j\left\langle(b+b^{\dagger})a_j\right\rangle \nonumber \\
& +E_j+\delta_{j1}E_m e^{-i\delta_{m}t},\label{eq:semiclassical 1}\\
\dot{\beta}=&(-\gamma_m-i\omega_m)\beta + \omega_{m}\frac{\beta_{\rm NL}}{3}e^{-i\varphi}\left\langle p_{\varphi}^3\right\rangle +i\sum_{j=1,2}   g_j \left\langle a_j^\dagger a_j \right\rangle ,
\label{eq:semiclassical_2}
\end{align}
\end{subequations}
where we have defined the optical and mechanical expectation values $\alpha_j = \langle a_j \rangle$ and $\beta= \langle b \rangle$, and the mechanical observable $p_{\varphi}=-i (b e^{i\varphi}-b^\dagger e^{-i\varphi})$ for compactness. Eqs.~(\ref{eq:semiclassical}) do not form a closed set of equations due to the presence of the second and third order moments $\left\langle(b+b^{\dagger})a_j\right\rangle$, $\left\langle p_{\varphi}^3\right\rangle$, and $\left\langle a_j^\dagger a_j \right\rangle$. The latter are independent variables from $\alpha_j$ and $\beta$, and their evolution equations involve all the higher order moments, yielding an infinite hierarchy of equations which cannot be exactly solved in general.

However one can derive a self-consistent treatment which is valid in a wide parameter region of optomechanical systems, which is reminiscent of the widely used Bogoliubov approximation \cite{Bogol47}, where a generic operator is separated into expectation value plus quantum fluctuations. Then, taking into account the first nonlinear terms one obtains the back-reaction corrections to the mean field dynamics induced by quantum fluctuations. 
In fact, we rewrite each Heisenberg representation operator as the sum of its expectation value with the corresponding, zero-mean, quantum fluctuation operator, that is,
\begin{eqnarray}
  a_j(t)& = & \alpha_j(t)+\delta a_j (t),\label{sep1}\\
  b(t) & = &  \beta(t) +\delta b(t),\label{sep2}
\end{eqnarray}
so that Eqs.~(\ref{eq:semiclassical}) can be rewritten as
\begin{subequations}\label{eq:semiclassicalbis}
\begin{align}
&\dot{\alpha_j}=(-\kappa_j+i\Delta^{(0)}_j)\alpha_j+ig_j(\beta+\beta^*)\alpha_j +E_j+\delta_{j1}E_m e^{-i\delta_{m}t}\nonumber \\
& +ig_j\left\langle(\delta b+\delta b^{\dagger})\delta a_j\right\rangle, \label{eq:semiclassicalbis 1}\\
&\dot{\beta}=(-\gamma_m-i\omega_m)\beta + \omega_{m}\frac{\beta_{\rm NL}}{3}e^{-i\varphi}\pi_{\varphi}^3 +i\sum_{j=1,2}   g_j |\alpha_j|^2 \nonumber \\
&+\omega_{m}\frac{\beta_{\rm NL}}{3}e^{-i\varphi}\left\langle \delta p_{\varphi}^3\right\rangle +\omega_{m}\beta_{\rm NL}\pi_{\varphi}e^{-i\varphi}\left\langle \delta p_{\varphi}^2 \right\rangle \nonumber \\
&+i\sum_{j=1,2}   g_j \left\langle \delta a_j^\dagger \delta a_j \right\rangle ,
\label{eq:semiclassicalbis_2}
\end{align}
\end{subequations}
where we have introduced the shorthand notation $\pi_{\varphi}=\langle p_{\varphi}\rangle =-i(\beta e^{i\varphi}-\beta^* e^{-i\varphi})=2 {\rm Im}(\beta e^{i\varphi})$. This latter set of equation is equivalent to Eqs.~(\ref{eq:semiclassical}) but it explicitly shows how the dynamics of the average values $\alpha_j$ and $\beta$ is affected by the covariances $\langle(\delta b+\delta b^{\dagger})\delta a_j\rangle$, $ \langle \delta a_j^\dagger \delta a_j \rangle $, $\langle \delta p_{\varphi}^2 \rangle$, and by $\langle \delta p_{\varphi}^3 \rangle $. Moreover it suggests an approximated treatment which can be applied in a very large parameter regime.

In fact, in many optomechanical systems, nonlinearities are quite small, because the single-photon optomechanical couplings $g_j$ are typically orders of magnitude smaller than the other relevant rates $\kappa_j$ and $\omega_m$, and mechanical nonlinearities are typically very small too, because it is often $\beta_{\rm NL} \ll 1$. A linear system which is affected by Gaussian noises possesses fluctuations which maintain the Gaussian properties, and therefore, due to the smallness of nonlinearities of typical optomechanical systems, it is reasonable to assume that the quantum fluctuations still maintain a Gaussian dynamics. This implies that $\langle \delta p_{\varphi}^3\rangle \simeq 0$, and that all the dynamical and statistical properties of the system can be expressed in terms of the first order expectation values and of the second-order covariance matrix. The dynamics of these quantities can be then described by two interconnected sets of coupled equations, \emph{the set of nonlinear equations for the expectation values}
\begin{subequations}\label{eq:semiclassicaltris}
\begin{align}
&\dot{\alpha_j}=(-\kappa_j+i\Delta^{(0)}_j)\alpha_j+ig_j(\beta+\beta^*)\alpha_j +E_j\nonumber \\
& +\delta_{j1}E_m e^{-i\delta_{m}t}+ig_j\left\langle(\delta b+\delta b^{\dagger})\delta a_j\right\rangle, \label{eq:semiclassicaltris 1}\\
&\dot{\beta}=(-\gamma_m-i\omega_m)\beta + \omega_{m}\frac{\beta_{\rm NL}}{3}e^{-i\varphi}\pi_{\varphi}^3 +i\sum_{j=1,2}   g_j |\alpha_j|^2 \nonumber \\
&+\omega_{m}\beta_{\rm NL}\pi_{\varphi}e^{-i\varphi}\left\langle \delta p_{\varphi}^2 \right\rangle+i\sum_{j=1,2}   g_j \left\langle \delta a_j^\dagger \delta a_j \right\rangle ,
\label{eq:semiclassicaltris_2}
\end{align}
\end{subequations}
and the \emph{linearized quantum Heisenberg-Langevin equations for the fluctuation operators}
\begin{subequations}\label{eq:quantumfluct}
\begin{align}
&\delta\dot{a}_j=\left[-\kappa_j+i\Delta^{(0)}_j+2ig_j{\rm Re}(\beta)\right]\delta a_j\nonumber \\
&+ig_j\alpha_j(\delta b+\delta b^\dagger)+\sqrt{2\kappa_{j,in}}a_{j,in}+\sqrt{2\kappa_{j,ex}}a_{j,ex}, \label{eq:quantumfluct1}\\
&\delta\dot{b}=(-\gamma_m-i\omega_m) \delta b\nonumber +i\sum_{j=1,2}   g_j(\alpha_j \delta a_j^\dagger +\alpha_j^* \delta a_j)\\
& -4 i \omega_{m}\beta_{\rm NL}\left[{\rm Im}(\beta e^{i\varphi})\right]^2 (\delta b -\delta b^{\dagger}e^{-2i\varphi}) +\sqrt{2\gamma_m}b_{in}.
\label{eq:quantumfluct2}
\end{align}
\end{subequations}
Eqs.~(\ref{eq:semiclassicaltris}) are driven not only by the pump and probe fields, but also by three covariance elements, which are determined by the solution of Eqs.~(\ref{eq:quantumfluct}), in which, in turn, the solutions $\alpha_j(t)$ and $\beta(t)$ of Eqs.~(\ref{eq:semiclassicaltris}) enter as time-dependent coefficients. The two sets can be solved through iterations. One first solves Eqs.~(\ref{eq:semiclassicaltris}) for the average values taking at the first round the initial condition values for the unknown covariances $\langle(\delta b+\delta b^{\dagger})\delta a_j\rangle$, $ \langle \delta a_j^\dagger \delta a_j \rangle $, $\langle \delta p_{\varphi}^2 \rangle$. Then one inserts this solution into Eqs.~(\ref{eq:quantumfluct}) which can be solved, providing therefore new input values for the covariances within Eqs.~(\ref{eq:semiclassicaltris}), which are then solved again and so on. The second-order covariances can be easily obtained from Eqs.~(\ref{eq:quantumfluct}) in the following way. One rewrites them in matrix form as
\begin{equation}\label{compact1}
\dfrac{d}{dt}u=Su+\xi,
\end{equation}
where $u=(\delta a_1,\delta a^{\dagger}_1,\delta a_2,\delta a^{\dagger}_2,\delta b,\delta b^{\dagger})^T$ is the vector of variables, and
$\xi=(\nu_1,\nu_1^{\dagger},\nu_2,\nu_2^{\dagger},\sqrt{2\gamma_m}b_{in},\sqrt{2\gamma_m}b_{in}^{\dagger})^T$ is the vector of noises, where $\nu_j=\sqrt{2\kappa_{j,in}}a_{j,in}+\sqrt{2\kappa_{j,ex}}a_{j,ex}$. As a consequence, the time-dependent matrix of coefficients $S$ is given by
\begin{widetext}
\begin{equation}
\begin{split}
S=
\begin{pmatrix}
-\kappa_1+i\Delta^{\rm eff}_1 & 0 & 0 & 0 & ig_1\alpha_1 & ig_1\alpha_1\\
0 & -\kappa_1-i\Delta^{\rm eff}_1 & 0 & 0 & -ig_1\alpha_1 & -ig_1\alpha_1\\
0 & 0 & -\kappa_2+i\Delta^{\rm eff}_2 & 0 & ig_2\alpha_2 & ig_2\alpha_2\\
0 & 0 & 0 & -\kappa_2-i\Delta^{\rm eff}_2 & -ig_2\alpha_2 & -ig_2\alpha_2\\
ig_1\alpha_1^* & ig_1\alpha_1 & ig_2\alpha_2^* & ig_2\alpha_2 & -\gamma_m-i\omega_m' &  4 i e^{-2i\varphi}\omega_{m}\beta_{\rm NL}\left[{\rm Im}(\beta e^{i\varphi})\right]^2  \\
-ig_1\alpha_1^* & -ig_1\alpha_1 & -ig_2\alpha_2^* & -ig_2\alpha_2 & -4 i e^{2i\varphi}\omega_{m}\beta_{\rm NL}\left[{\rm Im}(\beta e^{i\varphi})\right]^2 &  -\gamma_m+i\omega_m'
\end{pmatrix},
\end{split}
\label{eq:iamgHamilton}
\end{equation}
where $\Delta^{\rm eff}_j= \Delta^{(0)}_j+2g_j {\rm Re}(\beta)$, $\omega_m'= \omega_m(1+4\beta_{\rm NL}\left[{\rm Im}(\beta e^{i\varphi})\right]^2) $.
From the definition of the covariance matrix $C_{ij}(t)=\langle u_i(t)u_j(t)+u_j(t)u_i(t)\rangle/2$, and the correlation function of the noise terms, one gets the following deterministic equation for the matrix $C$
\begin{equation}
\frac{dC}{dt}=SC+CS^{T}+N,
\label{compact2}
\end{equation}
where $N$ is the diffusion matrix
\begin{equation}
\begin{split}
N=
\begin{pmatrix}
0 & \kappa_1 (2n_1+1) & 0 & 0 & 0 & 0\\
\kappa_1 (2n_1+1) & 0 & 0 & 0 & 0 & 0 \\
0 & 0 & 0 & \kappa_2 (2n_2+1) & 0 & 0 \\
0 & 0 & \kappa_1 (2n_2+1) & 0 & 0 & 0 \\
0 & 0 & 0 & 0 & 0 & \gamma_m (2n_b+1) \\
0 & 0 & 0 & 0 & \gamma_m (2n_b+1) & 0
\end{pmatrix}.
\end{split}
\label{eq:noise}
\end{equation}
\end{widetext}
The mean thermal photon number of the pump and probe cavity modes, $n_j$, $j=1,2$, can be taken equal to zero, because $n_j=[\exp(\hbar\omega_{cj}/k_bT)-1]^{-1} \simeq 0$ at optical frequencies. On the contrary, the thermal equilibrum occupancy of the mechanical resonator, $n_b=[\exp(\hbar\omega_{m}/k_bT)-1]^{-1}$ is nonzero and may be very large even at cryogenic temperatures, for mechanical resonators in the MHz regime. 

The iterative solution of the two sets of deterministic equations, Eqs.~(\ref{eq:semiclassicaltris}) and Eq.~(\ref{compact2}) provides an approximate but fast and effective solution method. With few iterations it reproduces satisfactorily the behavior of the solution of the full Langevin equations of Eqs.~(\ref{eq:quantum langevin}) averaged over a sufficiently large number of random trajectories, whenever the single-photon optomechanical coupling and the mechanical nonlinearity are small and unable to induce appreciable non-Gaussian statistics.

We notice that the need of iterations comes only from the presence of the covariances $\langle(\delta b+\delta b^{\dagger})\delta a_j\rangle$, $ \langle \delta a_j^\dagger \delta a_j \rangle $, $\langle \delta p_{\varphi}^2 \rangle$ in Eqs.~(\ref{eq:semiclassicaltris}). If these covariance matrix elements are negligible at all times, Eqs.~(\ref{eq:semiclassicaltris}) can be immediately solved and their solution for the expectation values can be inserted within Eq.~(\ref{compact2}) to get also the time evolution of the covariances. In most optomechanical experiments the intracavity mean photon number is large, $|\alpha_j|^2 \gg \langle \delta a_j^\dagger \delta a_j \rangle \sim 1$, implying that the first two covariances, $\langle(\delta b+\delta b^{\dagger})\delta a_j\rangle$, $ \langle \delta a_j^\dagger \delta a_j \rangle $, are often negligible at all times. The mechanical variance $\langle \delta p_{\varphi}^2 \rangle$ is instead different, it is typically non-negligible at the beginning, since it is equal to the thermal equilibrium value $2 n_b+1$, and it can be neglected in Eqs.~(\ref{eq:semiclassicaltris}) only if the mechanical nonlinearity is extremely small, or when we are close to the quantum regime, $\langle \delta p_{\varphi}^2 \rangle \sim 1$.

\section{Preparation of a large amplitude mechanical coherent state and stationary measurement of nonlinearity}\label{stationary}

We now apply the approach of the previous Section to describe the preparation of a large-amplitude stationary coherent state of the mechanical resonator. We then discuss how such a state can be used to provide a high-sensitive estimation of the mechanical nonlinearity $\beta_{\rm NL}$. The typical initial condition is a factorized state with the vacuum state for the pump and probe cavity modes, which are then excited by the driving lasers to large amplitudes $\alpha_j$, and an initial thermal state for the mechanical resonator. The target stationary coherent state is achieved by realizing simultaneously ground state cooling and coherent excitation of the mechanical resonator. This state is generated by means of the pump and its modulation at $\delta_m$, when the pump is red detuned with respect to the cavity, $\Delta^{\rm eff}_1 \simeq -\omega_m$ in order to realize sideband cooling \cite{Aspelmeyer2014,Genes2009}, and the modulation is quasi-resonant with the cavity, $\delta_m \sim \omega_m$. In this way, the beats between the two tones are able to coherently excite the mechanical resonator to a large amplitude. As explained in the previous Section, we describe the dynamics neglecting the two covariance terms $\langle(\delta b+\delta b^{\dagger})\delta a_j\rangle$, $ \langle \delta a_j^\dagger \delta a_j \rangle $ in Eqs.~(\ref{eq:semiclassicaltris}), and then solving the latter together with Eq.~(\ref{compact2}) for the quantum fluctuations described by the covariance matrix elements. 

It is convenient to re-express the coherent motion of the resonator as
\begin{equation}\label{parambeta}
  \beta(t) = \beta_0 + A_b(t) e ^{-i\delta_m t},
\end{equation}
i.e., as a constant term $\beta_0$ and a term oscillating at the driving frequency $\delta_m$, so that $A_b(t)$ is a slowly varying amplitude, which changes slowly due to the damping and to the nonlinear terms. Inserting Eq.~(\ref{parambeta}) into Eq.~\eqref{eq:semiclassicaltris 1}, and solving it, formally, by neglecting the transient term related to the initial values $\alpha_j(0)$, we have
\begin{equation}
\begin{split}
\alpha_j(t)=&\int_0^tdt'\left\{e^{\mathcal{L}_j(t-t')}[E_j+\delta_{j1}E_m e^{-i\delta_{m}t'}]\right. \\
& \left. \times \exp\left[2i g_j \int_{t'}^t dt'' \vert A_b(t'')\vert\cos(\delta_m t''-\theta)\right] \right\},
\end{split}
\label{eq:cavity field1}
\end{equation}
where $\mathcal{L}_j=i [\Delta_j^{(0)}+ g_j(\beta_{0}+\beta^{*}_{0})]-\kappa_j$, and we have rewritten the complex amplitude as $A_b(t)=\vert A_b(t)\vert e^{i\theta}$.
The amplitude $A_b(t)$ is much slower than the fast oscillations at $\delta_m$ and one can treat it as a constant in the integral over $t''$ in Eq.~(\ref{eq:cavity field1}). Performing explicitly this integral one gets
\begin{equation}
\alpha_j(t)=e^{i\psi_j(t)}\int_0^tdt' e^{\mathcal{L}_j(t-t')}[E_j+\delta_{j1}E_m e^{-i\delta_{m}t'}]e^{-i\psi_j(t')},
\label{eq:cavity field2}
\end{equation}
where $\psi_j(t)=\xi_j\sin(\delta_m t-\theta)$, with $\xi_j=2g_j\vert A_b\vert /\delta_m$.
We then use the Jacobi-Anger expansion for the $e^{-i\psi_j(t')}$ factor within the integral, i.e., $e^{-i\xi_j\sin\phi}=\sum_n J_n(-\xi_j)e^{i\phi n}$, ($\phi = \delta_m t'-\theta$ and $J_n$ is the $n$-th Bessel function of the first kind), and after neglecting a quickly decaying term proportional to $e^{\mathcal{L}_j t}$, we finally get
\begin{eqnarray}
&&\alpha_j(t)  =  E_j e^{i\psi_j(t)}\sum_{n=-\infty}^{\infty}\dfrac{J_n\left(-\xi_j\right)e^{i(\delta_m t-\theta)n}}{i\delta_m n-\mathcal{L}_j}  \\
&&+ \delta_{j1}E_m e^{i\psi_j(t)}\sum_{n=-\infty}^{\infty}\dfrac{J_n\left(-\xi_j\right)e^{i(\delta_m t-\theta)n}e^{-i\delta_m t}}{i\delta_m (n-1)-\mathcal{L}_j}, \nonumber
\label{eq:cavity field solution}
\end{eqnarray}
which, shifting the index of the sum in the second term, can be rewritten as
\begin{eqnarray}
&& \alpha_j(t)  =  e^{i\psi_j(t)}\sum_{n=-\infty}^{\infty}\dfrac{e^{i(\delta_m t-\theta)n}}{i\delta_m n-\mathcal{L}_j} \nonumber \\
&& \times \left[E_j J_n\left(-\xi_j\right)+\delta_{j1}E_m J_{n+1}\left(-\xi_j\right) e^{-i\theta}\right].
\label{eq:cavity field solution2}
\end{eqnarray}
We have to insert this formal solution into Eq.~(\ref{eq:semiclassicaltris_2}) for the dynamics of the mechanical oscillator, and therefore we need the modulus squared of this latter expression, which reads
\begin{eqnarray}
&& \vert \alpha_j(t)\vert ^2  =  \sum_{n,n'=-\infty}^{\infty}\dfrac{e^{i(\delta_m t-\theta)(n-n')}}{(i\delta_m n-\mathcal{L}_j)(-i\delta_m n'-\mathcal{L}_j^*)} \nonumber\\
&& \times \left[E_j J_n\left(-\xi_j\right)+\delta_{j1}E_m J_{n+1}\left(-\xi_j\right) e^{-i\theta}\right] \nonumber \\
&& \times \left[E_j J_{n'}\left(-\xi_j\right)+\delta_{j1}E_m J_{n'+1}\left(-\xi_j\right) e^{i\theta}\right].  \label{doublesum}
\end{eqnarray}
This latter expression has to be inserted into Eq.~(\ref{eq:semiclassicaltris_2}) together with Eq.~(\ref{parambeta}); the equation for the constant shift $\beta_0$
can be obtained by considering only the non-oscillating terms, while the equation for the slowly varying amplitude $A_b(t)$ can be obtained by considering only the quasi-resonant terms oscillating at $\delta_m$. In fact, all the other terms oscillate at different harmonics and provide a negligible effect. For $\beta_0$ we get
\begin{eqnarray}
&&0=(-\gamma_m-i\omega_m)\beta_0 +i \omega_{m}\frac{\beta_{\rm NL}}{3} e^{-i\varphi}(\beta_0 e^{i\varphi}-\beta_0^* e^{-i\varphi})^3 \nonumber \\
&&-i \omega_{m}\beta_{\rm NL}(\beta_0 -\beta_0^* e^{-2i\varphi})\left(2|A_b|^2+\langle \delta p_{\varphi}^2 \rangle\right) \label{eq:zero-order steady} \\
&& +i\sum_{j=1,2}   g_j \sum_{n=-\infty}^{\infty}\frac{\left|E_j J_n\left(-\xi_j\right)+\delta_{j1}E_m J_{n+1}\left(-\xi_j\right) e^{-i\theta}\right|^2}{|i\delta_m n-\mathcal{L}_j|^2}. \nonumber
\end{eqnarray}
Eq.~(\ref{eq:zero-order steady}) cannot be easily used to determine the value of $\beta_{0}$ because of the presence of terms depending upon the slowly varying variables $A_b(t)$ and $\langle \delta p_{\varphi}^2 \rangle$. On the other hand, this static oscillator shift determines the effective cavity detunings
\begin{equation}
 \Delta_{j}=\Delta_{j}^{(0)} + g_j(\beta_{0}+\beta^{*}_{0}),
\end{equation}
which is the actual parameter controlled in an experiment with the Pound-Drever-Hall (PDH) \cite{pdh} locking circuit. As a consequence, $\mathcal{L}_j=i\Delta_{j}-\kappa_j$ become given known parameters, and one expects that Eq.~(\ref{eq:zero-order steady}) is self-consistently satisfied as soon as $A_b(t)$ reaches its stationary value.

However, if the nonlinearities are not too large, a simple approximate expression for $\beta_0$ can be obtained by neglecting the mechanical nonlinear terms proportional to $\beta_{\rm NL}$, taking the zero-order term in $\xi_j$ in the sum over $n$, and assuming that $E_m \ll E_1$. One gets
\begin{equation}\label{approxbeta0}
  \beta_0 \simeq \frac{\omega_m + i\gamma_m}{\omega_m^2+\gamma_m^2}\sum_{j=1,2}\frac{g_j E_j^2}{\kappa_j^2+\Delta_j^2}.
\end{equation}

As already mentioned above, the equation for the slowly varying coherent state amplitude $A_b(t)$ is obtained by considering in Eq.~(\ref{eq:semiclassicaltris_2}) only the terms behaving as $e^{-i\delta_m t}$. These are the only quasi-resonant terms when one chooses $\delta_m \sim \omega_m$, and this means considering only two terms in the expansion of the mechanical nonlinearity, and keeping only the terms with $n-n'=-1$ in the double sum of Eq.~(\ref{doublesum}). One gets
\begin{eqnarray}
&&\dot{A}_b(t)=\left(-\gamma_m-i\Delta_m\right) A_b(t)-i\beta_{\rm NL}\omega_m |A_b(t)|^2 A_b(t)  \nonumber \\
&&+i\sum_{j=1,2} g_j \sum_{n=-\infty}^{\infty}\dfrac{\left[e^{i\theta}E_j J_n\left(-\xi_j\right)+\delta_{j1}E_m J_{n+1}\left(-\xi_j\right) \right]}{(i\delta_m n-\mathcal{L}_j)[-i\delta_m (n+1)-\mathcal{L}_j^*]} \nonumber \\
&& \times \left[E_j J_{n+1}\left(-\xi_j\right)+\delta_{j1}E_m J_{n+2}\left(-\xi_j\right) e^{i\theta}\right],
\label{eq:first-order solution}
\end{eqnarray}
where
\begin{equation}\label{detmech}
\Delta_m =\omega_m-\delta_m+\beta_{\rm NL}\omega_m\left\{4\left[{\rm Im}\left(\beta_0 e^{i\varphi}\right)\right]^2+\langle p_{\varphi}^2\rangle\right\}.
\end{equation}
Eq.~(\ref{eq:first-order solution}) is the relevant equation determining the evolution of the amplitude of the target mechanical coherent state, $A_b(t)$. It includes two nonlinear terms, the mechanical nonlinearity given by $\beta_{NL}$, and the one associated with the radiation pressure force and expressed by the sum over Bessel functions.

\subsection{Series expansion of the radiation pressure terms and stationary estimation of the nonlinearity}

In order to understand the effect of the radiation pressure nonlinearity we will develop the sum terms in Eq.~(\ref{eq:first-order solution}) in series of powers of $\xi_j$. This is justified for a wide range of values of $|A_b|$, because in most cavity optomechanical systems $g_j/\delta_m \ll 1$. We will stop at third order in $\xi_j$ since the mechanical nonlinear term is also at third order in $A_b$. We use the fact that for $x \to 0$, $J_n(x) \sim x^n/n!2^n$ and $J_{-n}(x) = (-1)^n J_n(x)$, so that the radiation pressure contribution at third order can be written as
\begin{equation}\label{radpressgen}
  F_{\rm RP}=c_0 +c_1 A_b(t)+c_2 A_b(t)^2+c_{2m}|A_b(t)|^2+c_3 A_b(t) |A_b(t)|^2,
\end{equation}
where the $c_i$ are complex constants. After long but straightforward calculations, one gets
\begin{eqnarray}
  c_0 &=& \frac{i g_1 E_1 E_m}{(\kappa_1+i\Delta_1)[\kappa_1-i(\Delta_1+\delta_m)]}, \\
  c_{2m} \hspace{-5pt}&=& \hspace{-5pt} \frac{-2i g_1^3 E_1 E_m[i(\Delta_1+\delta_m)-\kappa_1]}{\delta_m^2(i\Delta_1-\kappa_1)[i(\Delta_1+2\delta_m)\hspace{-3pt}-\hspace{-3pt}\kappa_1][i(\Delta_1+\delta_m)\hspace{-3pt}+\hspace{-3pt}\kappa_1]}, \\
  c_{2} &=& \frac{c_{2m}}{2}-\frac{g_1^2}{\delta_m^2}c_0, \\
  c_1  &=& i\sum_{j=1,2}\frac{g_j^2  E_j^2}{\delta_m}\left[\frac{1}{(\kappa_j+i\Delta_j)[\kappa_j-i(\delta_m+\Delta_j)]}\right. \nonumber \\
  &&\left.-\frac{1}{(\kappa_j-i\Delta_j)[\kappa_j-i(\delta_m-\Delta_j)]}\right], \\
   c_3  &=& i\sum_{j=1,2}\frac{g_j^4  E_j^2}{2\delta_m^3}\left[\frac{1}{[\kappa_j+i(\delta_m+\Delta_j)][\kappa_j-i(2\delta_m+\Delta_j)]}\right. \nonumber \\
  &&\left.-\frac{1}{[\kappa_j+i(\delta_m-\Delta_j)][\kappa_j-i(2\delta_m-\Delta_j)]}\right],
\end{eqnarray}
where for simplicity we have neglected a contribution proportional to $E_m^2$ in the expression for $c_1$ and $c_3$, which is negligible as long as $E_m \ll E_1$.

Using Eq. (\ref{radpressgen}), Eq.~(\ref{eq:first-order solution}) at third order reads
\begin{eqnarray}
&&\dot{A}_b(t)=\left(-\gamma_m^{\rm eff}-i\Delta_m^{\rm eff}\right) A_b(t)+c_0 +c_2 A_b(t)^2 \nonumber \\
&&+c_{2m}|A_b(t)|^2+\left(c_3-i\beta_{\rm NL}\omega_m \right) |A_b(t)|^2 A_b(t),
\label{eq:first-order solution_appr}
\end{eqnarray}
where the effective parameters
\begin{eqnarray}
  \gamma_m^{\rm eff} &=& \gamma_m-{\rm Re}\left(c_1\right), \label{effdamp}\\
  \Delta_m^{\rm eff} &=& \Delta_m-{\rm Im}\left(c_1\right), \label{effshift}
\end{eqnarray}
describe the usual modification of damping associated with sideband cooling and the usual optical spring effect, respectively \cite{Aspelmeyer2014}.

The stationary solution of Eq.~(\ref{eq:first-order solution_appr}) obtained setting $\dot{A}_b(t)=0$ provides the amplitude of the target stationary coherent state $A_b^{st}$. The corresponding covariance matrix elements of the reduced mechanical state are instead obtained from the stationary solution of Eq.~(\ref{compact2}). However, since we operate in a regime where standard sideband cooling is efficient, the position and momentum variances will be very close to the zero-point fluctuations of the quantum ground state, with a thermal occupancy $n_0 \ll 1$ and a purity ${\cal P}\sim 1$, where ${\cal P}= \left(2 \sqrt{|{\rm Det} C_b|}\right)^{-1}$, with $C_b$ the covariance matrix of the reduced state of the mechanical resonator~\cite{Borkje}. 

\begin{figure}[!b]
	\centering
	\includegraphics[width=.99\linewidth]{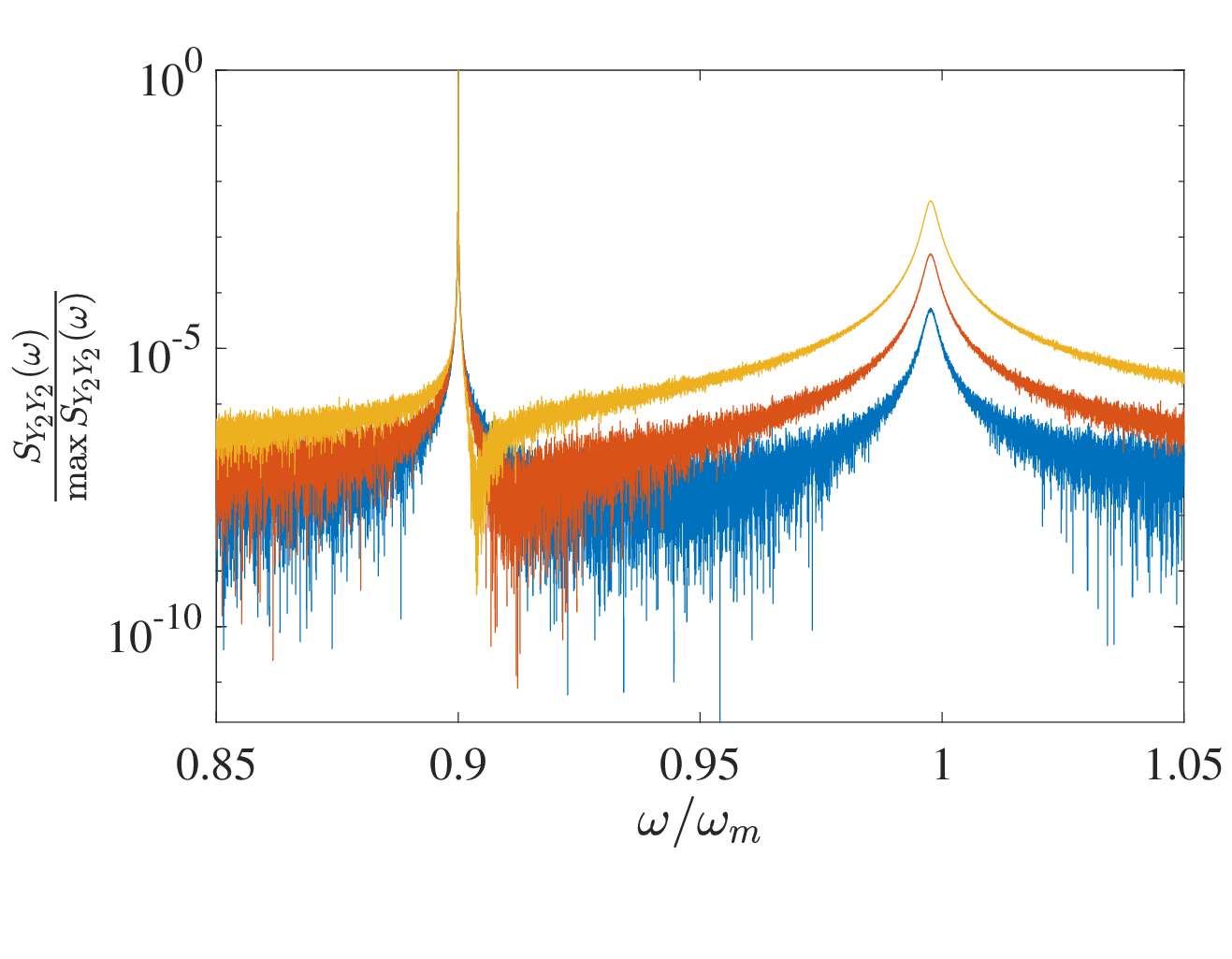}
	\caption{
Output probe spectrum versus $\omega/\omega_m$ for different values of the modulation power $P_m$: $P_m= 2 \times 10^{-8}$ W (blue), $P_m= 2 \times 10^{-7}$ W (red), $P_m= 2 \times 10^{-6}$ W (orange). Curves are obtained from the numerical solution of Eqs.~(\protect\ref{eq:quantum langevin}). The other parameters are: $\omega_m/2\pi =-\Delta_1/2\pi = 525$ kHz, $-\Delta_2/2\pi = 675$ kHz, $\kappa_1/2\pi = \kappa_2/2 \pi = 100$ kHz, $\kappa_{j,in}=\kappa_j/2$, j=1,2, $g_1/2\pi=g_2/2 \pi = 5$ Hz, $\beta_{NL}=10^{-14}$, $\gamma_m = \omega_m/Q_m$, with a mechanical quality factor $Q_m= 10^9$, $T = 100$ mK. We have assumed a carrier wavelength $\lambda = 1064$ nm, a cooling pump power $P_1 = 5 \times 10^{-5}$ W and a probe pump power $P_2 = 5 \times 10^{-8}$ W, corresponding respectively to $E_1/\omega_m = 3931 $, and $E_2/\omega_m = 124$. The achieved stationary mechanical occupancy is $n_0 = 0.015$, while the corresponding purity of the steady state is ${\cal P}=0.97$.  
	}
\label{fig:spectrapeak}
\end{figure}

At lowest order in the nonlinear coefficients $c_2$, $c_{2m}$, $c_3$, and $\beta_{NL}$, $A_b^{st}$ reads
\begin{eqnarray}\label{absta}
 && A_b^{st} \simeq \frac{c_0}{\Gamma^{\rm eff}}+\frac{c_2}{\Gamma^{\rm eff}}\left(\frac{c_0}{\Gamma^{\rm eff}}\right)^2 \\
 &&+\left|\frac{c_0}{\Gamma^{\rm eff}}\right|^2 \left[\frac{c_{2m}}{\Gamma^{\rm eff}}+\frac{\left(c_3-i\beta_{\rm NL}\omega_m \right)}{\Gamma^{\rm eff}} \frac{c_0}{\Gamma^{\rm eff}}\right], \nonumber
\end{eqnarray}
where $\Gamma^{\rm eff}=\gamma_m^{\rm eff}+i\Delta_m^{\rm eff}$. 
This expression naturally suggests a procedure for the estimation of the unknown nonlinear coefficient $\beta_{\rm NL}$, associated with the material nonlinearity or with the deformation parameter of quantum gravity theories. In fact, $A_b^{st}$ can be measured, for instance, from the height of the peak at the modulation frequency $\delta_m$ in the calibrated output spectrum of the probe beam (see Fig.~\ref{fig:spectrapeak}, or Fig. 4 of Ref.~\cite{bonaldi}). From the measured values of $A_b^{st}$ for different values of the various parameters (e.g., the detunings), and the knowledge of the other system parameters, one can in principle derive an estimate for $\beta_{\rm NL}$. However Eq.~(\ref{absta}) clearly shows that such an estimate may be hindered by the presence of the $c_3$ term associated with the third-order nonlinearity of the radiation pressure of the pump and probe modes.
More in detail, the imaginary part of $c_3$ can be roughly estimated to be of order $|{\rm Im}(c_1)| (g_j/\delta_m)^2$, where ${\rm Im}(c_1)$ is the mechanical frequency shift due to the optical spring [see Eq.~(\ref{effshift})], which is typically much smaller than $\delta_m \sim \omega_m$, say ${\rm Im}(c_1)\sim 10^{-3} \delta_m$. A typical value is $g_j/\delta_m \sim 10^{-5}$, and therefore the third-order radiation pressure nonlinearity is usually the dominant one, unless the mechanical nonlinearity or any effective nonlinearity associated with new physics is large enough, $\beta_{\rm NL} \gtrsim 10^{-13}$. This fact suggests that this stationary nonlinearity estimation scheme may be biased and not sensitive enough for probing new physics effects. 

In fact, a lower bound of $\beta_{NL} = 4.21 \times 10^{-21}$ has been reported in Ref.~\cite{Bawaj2015} using a nanogram-scale SiN membrane. This class of mechanical resonators is certainly suitable for the generation scheme of pure coherent states described here, because ground-state cooling has already been demonstrated with these membranes (see Ref.~\cite{Rossi2018} and references therein). 

Fig.~\ref{fig:spectrapeak} shows the homodyne output spectrum of the probe field 
\begin{eqnarray}
&&S_{Y_2 Y_2}^{{\rm out}}(\omega)=\frac{1}{2}\int_{-\infty}^{\infty} d\tau \left\langle Y_2^{{\rm out}}(t)Y_2^{{\rm out}}(t+\tau) \right. \nonumber \\
&& \left. +Y_2^{{\rm out}}(t+\tau)Y_2^{{\rm out}}(t)\right\rangle e^{i \omega \tau},  \label{probespectrum}
\end{eqnarray}
where $Y_2^{{\rm out}}(t) = -i [a_2^{{\rm out}}(t)-a_2^{{\rm out},\dagger}(t)]$, and $a_2^{{\rm out}}(t)=\sqrt{2 \kappa_{2,in}} a_2(t)-a_{2,in}(t)$, for different values of the power of the modulation tone at $\delta_m$. The quasi-resonant peak associated with pump modulation emerges from a broad Lorentzian-shaped pedestal associated with the ground-state cooling of the mechanical resonator. An accurate measurement of $A_b^{st}$ from the peak height requires a proper calibration of the vertical axis, and the calibration tone centered at $\omega/\omega_m = 0.9$ is visible in Fig.~\ref{fig:spectrapeak}. We underline that, for large enough coherent amplitudes such that $\xi_j \simeq g_j A_b^{st}/\omega_m > 1$, the relation between the output probe signal and the mechanical oscillation amplitude becomes highly nonlinear, and one has to adopt the calibration method described in Refs.~\cite{Piergentili2022,Piergentili2021}.


\section{Nonstationary dynamics: Turning off both the pump and its modulation}\label{pumpoff}

The above analysis suggests to look for alternative, dynamical, nonstationary schemes for a more sensitive estimation of $\beta_{NL}$. 
Nonetheless, the study of the stationary state of the previous Section provides the basis for understanding what happens if, after reaching the steady state, one turns off the pump driving $E_1 \to 0$, together with the modulation, $E_m \to 0$, and looks at the subsequent dynamics. The basic equations of motion in this second case are just a slight modification of
Eqs.~(\ref{eq:quantum langevin}),
\begin{subequations}\label{eq:quantum langevin-tran}
\begin{align}
\dot{a_j}=&(-\kappa_j+i\Delta^{(0)}_j)a_j+E_j e^{-\delta_{j1} t/\tau} \nonumber \\
&+ig_j(b+b^\dagger)a_j+\sqrt{2\kappa_{j,in}}a_{j,in}+\sqrt{2\kappa_{j,ex}}a_{j,ex}, \label{eq:quantum langevin1-tran}\\
\dot{b}=&(-\gamma_m-i\omega_m) b+ i e^{-i\varphi}\omega_{m}\frac{\beta_{\rm NL}}{3}(b e^{i\varphi}-b^\dagger e^{-i\varphi})^3 \nonumber \\
& +i\sum_{j=1,2}   g_ja_j^\dagger a_j+\sqrt{2\gamma_m}b_{in},
\label{eq:quantum langevin2-tran}
\end{align}
\end{subequations}
that is, we have eliminated the modulation $E_m$, and we have assumed that the pump driving has an exponential turning off dynamics with a decay time $\tau$.

\begin{figure}[!b]
	\centering
	\includegraphics[width=.75\linewidth]{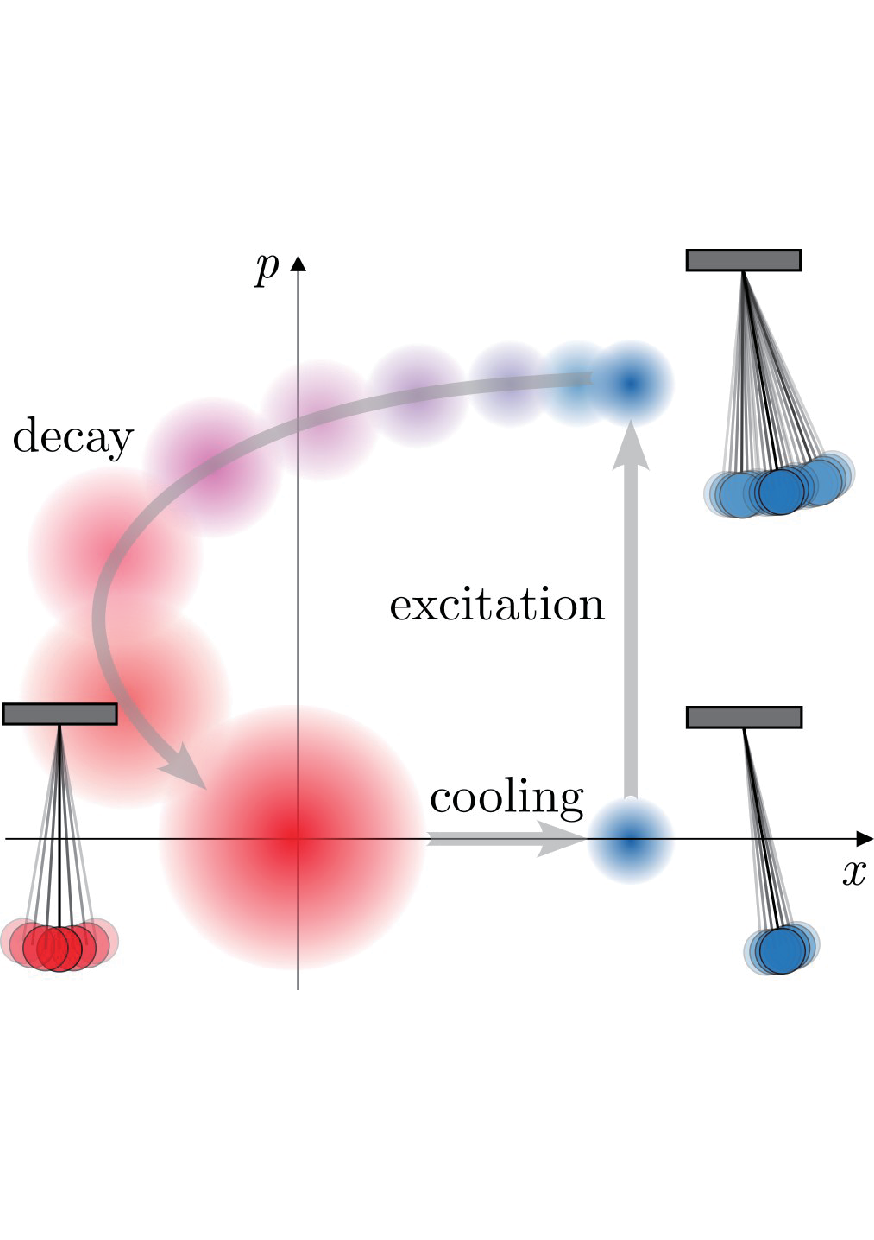}
	\caption{
Phase space description of the nonstationary dynamics which will be exploited for the estimation of $\beta_{NL}$. The combined action of the cooling pump and of its modulation prepares a large amplitude, almost pure, coherent state (upper right in the phase space), as described in Sec.~\protect\ref{stationary}. This state then decays to the thermal equilibrium state at the corresponding temperature of the reservoir.
	}
\label{fig:cartoon}
\end{figure}

A schematic description of the nonstationary dynamics in the phase space of the mechanical resonator, used for the estimation of $\beta_{NL}$, is shown in Fig.~\ref{fig:cartoon}. The prepared large amplitude coherent state decays to the thermal equilibrium state, and the weak nonlinearity is responsible for an amplitude-dependent frequency shift.

We follow the approach described above, i.e., we assume again the separation into a coherent classical part and quantum fluctuation of Eqs.~(\ref{sep1})-(\ref{sep2}). 

\subsection{The semiclassical amplitude dynamics}

The semiclassical dynamics of the cavity mode and mechanical amplitudes $\alpha_j(t)$ and $\beta(t)$ are obtained by adapting Eqs.~(\ref{eq:semiclassicaltris}) to the new conditions. As discussed in the previous Section we can neglect the second order covariances $\left\langle \delta a_j^\dagger \delta a_j \right\rangle$, $\left\langle(\delta b+\delta b^{\dagger})\delta a_j\right\rangle$. Moreover, since we start from an almost pure coherent-like state, one can also neglect $\left\langle \delta p_{\varphi}^2 \right\rangle$ which is now of the same order. One gets
\begin{subequations}\label{eq:classical_tran}
\begin{align}
&\dot{\alpha_j}=(-\kappa_j+i\Delta^{(0)}_j)\alpha_j+ig_j(\beta+\beta^*)\alpha_j +E_j e^{-\delta_{j1} t/\tau} , \label{eq:classical 1_tran}\\
&\dot{\beta}=(-\gamma_m-i\omega_m)\beta + \omega_{m}\frac{\beta_{\rm NL}}{3}e^{-i\varphi}\pi_{\varphi}^3 +i\sum_{j=1,2}   g_j |\alpha_j|^2. \label{eq:classical_2_tran}
\end{align}
\end{subequations}
We are interested in the non-stationary dynamics subsequent the pump switch-off and therefore the initial conditions for this set of equations are relevant, and are given by the steady state described in the previous Section. The mechanical mode starts in an almost pure coherent-like state, with very small effective occupancy $n_0$, which is not relevant for the dynamics of the amplitudes $\alpha_j(t)$ and $\beta(t)$. The two optical modes can be assumed to be each in a coherent state with amplitudes $\alpha_j(0)$ corresponding to the stationary values of the equations in the previous Section. Due to Eq.~(\ref{parambeta}), we have for the initial mechanical amplitude $\beta (0)= \beta_0 + A_b^{st} e^{-i\phi_0}$, where the constant shift $\beta_0$ is approximately given by Eq.~(\ref{approxbeta0}), $A_b^{st} \sim c_0/(\gamma_m^{\rm eff}+i \Delta_m^{\rm eff})$ if we neglect the nonlinear corrections in Eqs.~(\ref{eq:first-order solution_appr}) and~(\ref{absta}), and $\phi_0$ is a phase associated with the fast driven oscillations at $\delta_m$.

We have now to adjust the parametrization of $\beta(t)$ of Eq.~(\ref{parambeta}) to the new dynamical situation. Eq.~(\ref{approxbeta0}) and the absence of pump drive suggests that we have to assume a new constant term, $\beta_0' \neq \beta_0$, which can be significantly different from the first one. We have again to factorize a slowly varying amplitude $ A_b(t)$ from a fast oscillating term at the frequency $\omega_m$ (there is no external driving now). The time evolution of the modulus and phase of $ A_b(t)$ is affected by damping and by the nonlinearity we want to measure, and therefore it represents the main quantity of interest here. 

Due to the abrupt change of the radiation pressure force caused by the pump switch-off, and to the fact that we have to satisfy the given initial condition $\beta(0)$, we cannot exclude that the amplitude $A_b(t)$ has an initial fast transient of short duration. As a consequence, a convenient parametrization is
\begin{equation}\label{parambeta2}
  \beta(t) = \beta_0' + \left[A_{\rm fast}(t)+A_b(t)\right] e ^{-i\omega_m t},
\end{equation}
where $A_b(t)$ is slowly varying with respect to the fast timescales $\omega_m$ and $\kappa_j$, and $A_{\rm fast}(t) \gg A_b(t)$ holds true only for a short initial transient of duration $\tau_f$. The value of $A_{\rm fast}(0)$ is fixed by the initial condition $\beta (0)= \beta_0 + A_b^{st} e^{-i\phi_0}\simeq \beta_0'+A_{\rm fast}(0)$, so that $A_{\rm fast}(0)\simeq \beta_0 -\beta_0' + A_b^{st} e^{-i\phi_0}$.
Inserting Eq.~(\ref{parambeta2}) into Eq.~\eqref{eq:classical 1_tran}, and solving it formally, \emph{now including the transient term} related to the initial values $\alpha_j(0)$, we have
\begin{eqnarray}
&&\alpha_j(t)=\exp\left[2i g_j \int_{0}^t dt' \vert A_b(t')\vert\cos(\omega_m t'-\theta)\right] \nonumber \\
&& \times e^{\mathcal{L}_j t}e^{i\phi_j^{\rm fast}} \alpha_j(0)+\int_0^tdt'\left\{e^{\mathcal{L}_j(t-t')}E_j e^{-\delta_{j1} t'/\tau}\right. \nonumber \\
&& \left. \times \exp\left[2i g_j \int_{t'}^t dt'' \vert A_b(t'')\vert\cos(\omega_m t''-\theta)\right] \right\},  \label{eq:cavity field1_tran}
\end{eqnarray}
with $\phi_j^{\rm fast}=2 g_j {\rm Re}\left\{\int_0^{\tau_f} dt A_{\rm fast}(t)e ^{-i\omega_m t}\right\}\simeq 2 g_j \tau_f {\rm Re}\left\{A_{\rm fast}(0) \right\}$ the phase due to the fast transient amplitude $A_{\rm fast}(t)$ in the short time $\tau_f$ and which, because of that, gives a negligible contribution within the integral term in the second line of Eq.~(\ref{eq:cavity field1_tran}). Moreover, similarly to what has been done in the stationary case, $\mathcal{L}_j=i [\Delta_j^{(0)}+ g_j(\beta_{0}'+\beta^{*'}_{0})]-\kappa_j$, and we have again rewritten the slowly varying amplitude as $A_b(t)=\vert A_b(t)\vert e^{i\theta}$,
and treated it as a constant in the integrals in Eq.~(\ref{eq:cavity field1_tran}). Performing the integrals one gets
\begin{eqnarray}
&&\alpha_j(t)=e^{i\psi_j(t)}\left\{e^{\mathcal{L}_j t} \alpha_j(0)e^{i\phi_j^{\rm fast}-i\psi_j(0)}\right. \label{eq:cavity field2_tran}  \\
&&\left. +\int_0^tdt' e^{\mathcal{L}_j(t-t')}E_j e^{-\delta_{j1} t'/\tau}e^{-i\psi_j(t')}\right\}, \nonumber
\end{eqnarray}
where now $\psi_j(t)=\tilde{\xi}_j\sin(\omega_m t-\theta)$, with $\tilde{\xi}_j=2g_j\vert A_b\vert /\omega_m$.
Following the same step as in Sec. IV, we finally get
\begin{eqnarray}
&&\alpha_j(t)  =  e^{i\psi_j(t)}\left\{e^{\mathcal{L}_j t} \alpha_j(0)e^{i\phi_j^{\rm fast}-i\psi_j(0)} \right.\label{eq:cavity field solution_tran}\\
&&- \left.E_j e^{\mathcal{L}_j t}\sum_{n=-\infty}^{\infty}\dfrac{J_n\left(-\tilde{\xi}_j\right)e^{-i\theta n}}{i\omega_m n-\mathcal{L}_j-\delta_{j1}/\tau}\right.   \nonumber  \\
&& \left.+ E_j e^{-\delta_{j1} t/\tau}\sum_{n=-\infty}^{\infty}\dfrac{J_n\left(-\tilde{\xi}_j\right)e^{i(\omega_m t-\theta)n}}{i\omega_m n-\mathcal{L}_j-\delta_{j1}/\tau} \right\}.  \nonumber
\end{eqnarray}
This is a general and exact expression for the cavity field amplitudes, and in order to better understand it, we can separate it into a transient and a long-time term:
\begin{eqnarray}
&&\alpha_j(t)  =  e^{i\psi_j(t)}\left\{\alpha_j^{\rm trans}(t)\right. \label{eq:cavity field solution_tranrewri}  \\
&& \left.+ \delta_{j2}E_2 \sum_{n=-\infty}^{\infty}\dfrac{J_n\left(-\tilde{\xi}_2\right)e^{i(\omega_m t-\theta)n}}{i\omega_m n-\mathcal{L}_2} \right\},  \nonumber
\end{eqnarray}
where the initial transient term is
\begin{eqnarray}
&&\alpha_j^{\rm trans}(t)  =  e^{\mathcal{L}_j t}\left\{\right. \alpha_j(0)e^{i\phi_j^{\rm fast}-i\psi_j(0)} \label{eq:cavity field solution_tran_tran}\\
&&- \left. E_j \sum_{n=-\infty}^{\infty}\dfrac{J_n\left(-\tilde{\xi}_j\right)e^{-i\theta n}}{i\omega_m n-\mathcal{L}_j-\delta_{j1}/\tau}\right\}   \nonumber  \\
&& + \delta_{j1}E_1 e^{- t/\tau}\sum_{n=-\infty}^{\infty}\dfrac{J_n\left(-\tilde{\xi}_1\right)e^{i(\omega_m t-\theta)n}}{i\omega_m n-\mathcal{L}_1-1/\tau},  \nonumber
\end{eqnarray}
that is, the sum of a term exponentially decaying with rate $\kappa_j$ both for the pump and probe amplitude, and a term for only the pump mode, decaying with time $\tau$ associated with the non-instantaneous switch-off of the pump drive.

We have to insert this formal solution into Eq.~(\ref{eq:classical_2_tran}) for the dynamics of the mechanical oscillator, and therefore we need the modulus squared of Eq.~(\ref{eq:cavity field solution_tranrewri}), which is much more involved than the one of the stationary case,
\begin{eqnarray}
&& \vert \alpha_j(t)\vert ^2  =  \vert \alpha_j(t)^{\rm trans}\vert ^2 \label{doublesum_tran}\\
&&+\delta_{j2}\sum_{n,n'=-\infty}^{\infty}\dfrac{E_2^2 J_n\left(-\tilde{\xi}_2\right)J_{n'}\left(-\tilde{\xi}_2\right)e^{i(\omega_m t-\theta)(n-n')}}{(i\omega_m n-\mathcal{L}_2)(-i\omega_m n'-\mathcal{L}_2^*)} \nonumber\\
&& +\delta_{j2} \alpha_j^{\rm trans}(t) E_2 \sum_{n=-\infty}^{\infty}\dfrac{J_n\left(-\tilde{\xi}_2\right)e^{-i(\omega_m t-\theta)n}}{-i\omega_m n-\mathcal{L}_2^*} \nonumber \\
&& +\delta_{j2} \alpha_j^{\rm trans,*}(t) E_2 \sum_{n=-\infty}^{\infty}\dfrac{J_n\left(-\tilde{\xi}_2\right)e^{i(\omega_m t-\theta)n}}{i\omega_m n-\mathcal{L}_2} .  \nonumber
\end{eqnarray}
This radiation pressure term has to be inserted into Eq.~(\ref{eq:classical_2_tran}) and one has to use also the parametrization of Eq.~(\ref{parambeta2}) in order to derive effective equations for the constant shift $\beta_0'$ and the slowly varying amplitude $A_b(t)$, which is the main quantity of interest here.

Eq.~(\ref{doublesum_tran}) suggests that the resulting dynamical equation for $A_b(t)$ is much more involved than those derived for the stationary approach, Eq.~(\ref{eq:first-order solution}) and Eq.~(\ref{eq:first-order solution_appr}). We now show that this is not actually true. On the contrary, under realistic conditions, the dynamical decay of $A_b(t)$ obeys a simpler evolution equation, which is particularly suitable for the estimation of the nonlinear parameter $\beta_{NL}$. This important simplification is due to two facts.

\emph{(i) The fast transient cavity amplitudes $\alpha_j(t)^{\rm trans}$ influence only the fast transient term $A_{\rm fast}(t)$, which is therefore decoupled from the slow dynamics of $A_b(t)$.} In fact, since $\kappa_j \gg \gamma_m$, and when the pump is turned off not too slowly, say $\kappa_j \tau \sim 1$, $A_{\rm fast}(t)$ will decay to zero in a very short time $\tau_f$ compared to the typical timescales of $A_b(t)$, governed by the mechanical damping rate and the nonlinearities. As a consequence, the radiation pressure force affecting the constant shift $\beta_0'$ and $A_b(t)$ is given only by the second line of Eq.~(\ref{doublesum_tran}), corresponding to the stationary term associated with the probe mode only.

For the constant shift $\beta_0'$ we can repeat the same arguments described in Sec. IV (see Eqs.~(\ref{eq:zero-order steady})-(\ref{approxbeta0})), applied to the case when only the driving $E_2 \neq 0$.
We recall that $\beta_0'$ determines the effective probe mode detuning,
\begin{equation}
\Delta_{2}=\Delta_{2}^{(0)} + g_2(\beta_{0}'+\beta^{*'}_{0}),
\end{equation}
which is the actual parameter controlled in an experiment with the PDH locking circuit. As a consequence, $\mathcal{L}_2=i\Delta_{2}-\kappa_2$ becomes a given known parameter. Similarly to the derivation of Eq.~(\ref{approxbeta0}), a good approximate expression for $\beta_0'$ is given by
\begin{equation}\label{approxbeta0_tran}
  \beta_0' \simeq \frac{\omega_m + i\gamma_m}{\omega_m^2+\gamma_m^2}\frac{g_2 E_2^2}{\kappa_2^2+\Delta_2^2}.
\end{equation}
This shows that $\beta_0' \neq \beta_0$, implying that the probe detuning $\Delta_2$ undergoes a sudden, appreciable change when the pump is turned off. Such a change must be compensated as quickly as possible by the PDH control, in order to keep the probe driving and the cavity mode locked. 

\emph{(ii) The radiation pressure force due to the stationary probe field is exactly zero, at all orders, when the probe is kept at resonance $\Delta_2 =0$} (see also Ref.~\cite{Piergentili2022,Piergentilig0}).
In fact, the slowly varying amplitude $A_b(t)$ is driven only by the quasi-resonant terms behaving as $e^{-i\omega_m t}$ in Eq.~(\ref{eq:classical_2_tran}). This implies keeping only the terms with $n-n'=-1$ in the double sum of Eq.~(\ref{doublesum_tran}), that is, a probe radiation pressure force term
\begin{equation}
i g_2 E_2^2 e^{i\theta}\sum_{n=-\infty}^{\infty}\dfrac{J_n\left(-\tilde{\xi}_2\right)J_{n+1}\left(-\tilde{\xi}_2\right)}{(i\omega_m n-\mathcal{L}_2)[-i\omega_m (n+1)-\mathcal{L}_2^*]} .
\label{eq:first-order solution_tran}
\end{equation}
When $\Delta_2 = 0$, for every term $\bar{n}$ of this infinite sum, there is the term with $n+1=-\bar{n}$ which is its opposite, due to the fact that $J_{-n}(x) = (-1)^n J_n(x)$, i.e., Eq.~(\ref{eq:first-order solution_tran}) is zero at all orders. 

Therefore, we reach the important conclusion that a probe mode perfectly resonant with the cavity realizes a noninvasive detection of the mechanical mode and of its nonlinearity, without any backaction, as it occurs in a Michelson interferometer readout, used for example in Ref.~\cite{Bawaj2015}. Therefore, in the resonant probe case, the equation for the slowly varying mechanical amplitude is simply given by
\begin{equation}
\dot{A}_b(t)=\left(-\gamma_m-i\Delta_m'\right) A_b(t)-i\beta_{\rm NL}\omega_m |A_b(t)|^2 A_b(t),
\label{eq:first-order solution_tran2}
\end{equation}
where now [see also Eq.~(\ref{detmech})]
\begin{equation}\label{detmech2}
\Delta'_m =4\beta_{\rm NL}\omega_m\left[{\rm Im}\left(\beta'_0 e^{i\varphi}\right)\right]^2.
\end{equation}

\subsection{Nonlinearity estimation from the nonstationary dynamics of the slowly varying mechanical amplitude}

We now show how the nonstationary decay of $A_b(t)$ under the conditions detailed above can be used to provide a sensitive estimation of the mechanical nonlinearity (and deformation parameter) $\beta_{\rm NL}$. In fact, Eq.~(\ref{eq:first-order solution_tran2}) can be solved exactly: we rewrite
\begin{equation}\label{abtilde}
  A_b(t)=\tilde{A}_b(t)e^{-(\gamma_m+i \Delta_m')t},
\end{equation}
so that Eq.~(\ref{eq:first-order solution_tran2}) yields the simpler equation for $\tilde{A}_b(t)$
\begin{equation}\label{abtildeeq}
 \dot{\tilde{A}}_b(t)=-i\beta_{\rm NL}\omega_me^{-2\gamma_m t}\vert\tilde{A}_b(t)\vert^2 \tilde{A}_b(t).
\end{equation}
Rewriting it as an equation for modulus and phase, $ \tilde{A}_b(t)=\vert\tilde{A}_b(t)\vert e^{i\tilde{\theta}(t)}$, we easily see that the modulus is constant $\vert\tilde{A}_b(t)\vert = \vert\tilde{A}_b(0)\vert =\vert A_b(0)\vert $, and that
\begin{equation}\label{thetatilde}
  \dot{\tilde{\theta}}(t)= -\beta_{\rm NL}\omega_m e^{-2\gamma_m t}\vert A_b(0)\vert^2,
\end{equation}
giving the solution
\begin{equation}\label{thetatildesol}
  \tilde{\theta}(t)= \theta_0 -\beta_{\rm NL}\frac{\omega_m}{2\gamma_m} \vert A_b(0)\vert^2\left(1-e^{-2\gamma_m t}\right).
\end{equation}
Using Eq.~(\ref{abtilde}), we finally get
\begin{eqnarray}
&&  A_b(t)=A_b(0)e^{-(\gamma_m+i \Delta_m')t }\label{abtildesol}  \\
&& \times \exp\left\{-i\beta_{\rm NL}\frac{\omega_m}{2\gamma_m} \vert A_b(0)\vert^2\left(1-e^{-2\gamma_m t}\right)\right\},\nonumber
\end{eqnarray}
which means that the mechanical oscillator decays with rate $\gamma_m$, and with an amplitude-dependent instantaneous frequency
\begin{equation}\label{abtildesolinst}
  \omega_{\rm inst}= \omega_m+\Delta_m'-\dot{\tilde{\theta}}=\Delta_m'+\omega_m\left(1+\beta_{\rm NL}e^{-2\gamma_m t}\vert A_b(0)\vert^2\right).
\end{equation}
This expression analytically describes the estimation of the nonlinear deformation parameter $\beta_{NL}$ introduced in Ref.~\cite{Bawaj2015}, and then also adopted in Refs.~\cite{bushev,tobar}. In fact, by measuring the phase of $A_b(t)$, its derivative provides the instantaneous frequency $\omega_{\rm inst}$, and fitting the coefficient of $\omega_{\rm inst}$ versus $\vert A_b(0)\vert^2$, or versus $e^{-2\gamma_m t}\vert A_b(0)\vert^2$, one gets an estimation of $\beta_{\rm NL}$.

However, as discussed in Sec. III, we also have to consider the time evolution of the mechanical covariance matrix elements during the decay of the amplitude $A_b(t)$. First, these covariances evolve with negligible radiation pressure effects. In fact, the pump cavity mode rapidly decays to the vacuum state and remains uncoupled with the mechanical resonator. Moreover, as discussed in the previous subsection, the probe cavity mode is at resonance, and its backaction on the mechanical resonator is exactly zero. Under this condition, the probe output phase quadrature performs a real-time quantum nondemolition measurement of the mechanical resonator position~\cite{Giovannetti2001}. 

In contrast, the effect of mechanical nonlinearity cannot be neglected, even if the parameter $\beta_{NL}$ is very small, because it can be significantly amplified by the initial large amplitude of the coherent-like state generated (see the non-diagonal element of the matrix $S$ of Eq.~(\ref{eq:iamgHamilton}), $-4 i e^{2i\varphi}\omega_{m}\beta_{\rm NL}\left[{\rm Im}(\beta e^{i\varphi})\right]^2$). 

In fact, in the description of the thermalization process with the reservoir at temperature $T$, we can safely neglect fast-rotating terms and get an effective Kerr nonlinear term, $$H_{\rm Kerr} = \hbar \omega_{m}\beta_{\rm NL}(b^\dagger b)^2/2.$$ This term explains why the estimation of $\beta_{NL}$ is independent from the phase $\varphi$, and it is unable to distinguish a Duffing nonlinearity ($\varphi = \pi/2$) from an effective deformed commutator nonlinearity ($\varphi = 0$). The Kerr nonlinearity generates some squeezing of the mechanical state (see Ref.~\cite{Daniel,Kartner}), but it does not modify the time evolution of the mean phonon number $n_b(t) = \left\langle \delta b^{\dagger}(t) \delta b(t) \right\rangle $, which is still given by the standard damped thermal harmonic oscillator expression
\begin{equation}\label{eq:thermal}
  n_b(t) = n_0 e^{-2\gamma_m t} + n_{b} \left(1-e^{-2\gamma_m t}\right),   
\end{equation}
where $n_{b}$ is the equilibrium mechanical occupancy at temperature $T$~\cite{Daniel,Kartner}. The heating rate of the mechanical resonator is therefore given by the initial time derivative of this expression, that is, $2\gamma_m n_{b}$, which can be small for mechanical resonators with a good mechanical quality factor in a cryogenic environment. Therefore, if one can perform an efficient measurement of the mechanical frequency shift for a time shorter than $ 1/2\gamma_m n_{b}$, one can estimate the deformation parameter $\beta_{NL}$ in a regime dominated by quantum fluctuations only. 

\subsection{Numerical results for the estimation of the nonlinearity}

Now we numerically solve the quantum Langevin equations of the system in order to establish the sensitivity limits of the proposed scheme for the estimation of $\beta_{NL}$. We simulate the full-time evolution, i.e., the preparation of the large-amplitude coherent state described in Sec. IV, and its subsequent decay after turning off the pump drive together with its modulation tone, described by Eqs.~(\ref{eq:quantum langevin-tran}). For simplicity we have assumed an instantaneous turn-off, $\tau = 0$, in the simulations.

We have to include the effect of the PDH cavity locking system, which must keep the detunings of the pump and probe modes fixed, also during the abrupt modification due to the pump turn-off. This is done by modifying Eqs.~(\ref{eq:quantum langevin-tran}) in the following way
\begin{subequations}\label{eq:quantum langevin-tran-pdh}
\begin{align}
&\dot{a_j}=\left[-\kappa_j+i\Delta^{(0)}_j)+ig_j(b+b^\dagger-x_{\rm PDH})\right]a_j \label{eq:quantum langevin1-tran-pdh}\\
&+E_j e^{-\delta_{j1} t/\tau-i\delta_{j2}\phi \sin \Omega_c t} +\sqrt{2\kappa_{j,in}}a_{j,in}+\sqrt{2\kappa_{j,ex}}a_{j,ex}, \nonumber \\
&\dot{b}=(-\gamma_m-i\omega_m) b+ i e^{-i\varphi}\omega_{m}\frac{\beta_{\rm NL}}{3}(b e^{i\varphi}-b^\dagger e^{-i\varphi})^3 \nonumber \\
& +i\sum_{j=1,2}   g_ja_j^\dagger a_j+\sqrt{2\gamma_m}b_{in},
\label{eq:quantum langevin2-tran-pdh}
\end{align}
\end{subequations}
that is, by subtracting from the detunings the low-frequency dynamics of the mechanical resonator over a bandwidth $\tau_{\rm PDH}^{-1}$, $ig_j x_{\rm PDH}= (ig_j/\tau_{\rm PDH})\int_{t-\tau_{\rm PDH}}^t ds\left[b(s)+b^\dagger(s)\right]$. We have assumed $\tau_{\rm PDH}=5.25 \times 10^{-6}\gamma_m^{-1}$ in our simulations. This modification allows us to keep the probe mode always at resonance, which, as we have seen above, is crucial to avoid any radiation pressure effect. Moreover, we have explicitly added the calibration tone as a phase modulation of the probe, $\phi \sin \Omega_c t$, following the treatment of Refs.~\cite{Piergentili2022,Piergentili2021}.

\begin{figure}[!b]
	\centering
\includegraphics[width=.99\linewidth]{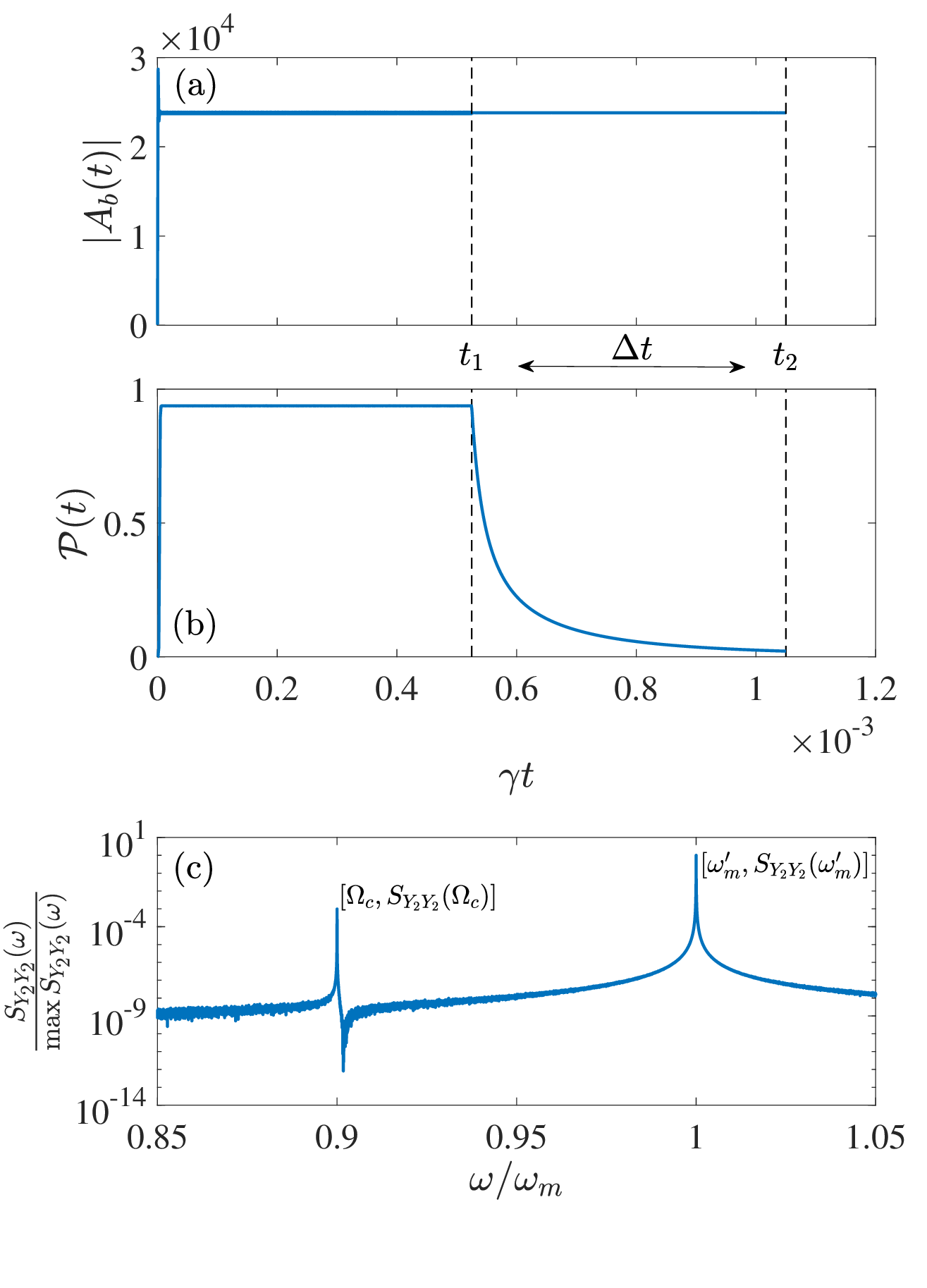}
\caption{Time evolution of the coherent amplitude $|A_b(t)|$ (a), and of the purity ${\cal P}(t)$ (b), of the mechanical reduced state. In the time interval from $t=0$ to $t=t_1$ a very pure large-amplitude coherent state is prepared. At $t=t_1$ the cooling pump and its modulation are turned off, and $|A_b(t)|$ decays with a slow timescale of order $\gamma_m^{-1}$, while ${\cal P}(t)$ decays with a faster timescale of order $(\gamma_m n_b)^{-1}$. In (c) we plot the output probe homodyne spectrum $S_{Y_2Y_2}(\omega)$ obtained via a FFT over the time interval $\Delta t = t_2-t_1 = 5.25 \times 10^{-4}\gamma_m^{-1}$ The calibration peak and the resonant signal peak which are used for the estimation of $\beta_{NL}$ are clearly visible. Curves are obtained by averaging over the numerical solution of Eqs.~(\protect\ref{eq:quantum langevin-tran-pdh}). The parameters are the same of Fig. \protect\ref{fig:spectrapeak}, and with $P_m = 2 \times 10^{-6}$ W, corresponding to $E_m/\omega_m = 786$.
	}
\label{fig1March}
\end{figure}

A typical time evolution obtained by averaging the trajectories obtained from the simulated Langevin equations is shown in Fig.~\ref{fig1March}(a) and Fig.~\ref{fig1March}(b), for the set of parameters given in the caption of Fig.~\ref{fig:spectrapeak}. The chosen set of parameters is very similar to that of the membrane-in-the-middle experiment of Ref.~\cite{bonaldi}, except that we have considered a higher mechanical quality factor ($Q = 10^9$), and a lower environmental temperature ($T=100$ mK). However, values comparable to these have been recently demonstrated in similar membrane-in-the-middle experiments~\cite{Rossi2018,marzioni}, and therefore the simulations performed here describe a situation achievable with current technology.

In the first time interval of duration $t_1 =5.25 \times 10^{-4} \gamma_m^{-1}$, the large--amplitude coherent state is prepared.
Fig.~\ref{fig1March}(a) shows the time evolution of its amplitude $|A_b(t)|$ which, starting from zero, in a very short time interval of order $1/\gamma_m^{\rm eff}$ not visible in the plot, reaches its stationary value $|A_b^{st}| \gg 1$. In the subsequent time interval $\Delta t = t_2-t_1 = 5.25 \times 10^{-4}\gamma_m^{-1}$, when the cooling pump and its modulation are turned off, the amplitude starts to decay very slowly (on a timescale of order $\gamma_m^{-1}$). Fig.~\ref{fig1March}(b) instead shows the time evolution of the purity ${\cal P}(t)$ of the mechanical resonator state. The purity starts from a very small value associated with the initial thermal state of the mechanical resonator and again, in a very short time interval, it reaches a steady state value very close to one. This shows that the pump and its modulation have generated an almost pure coherent state, as we have verified also from the stationary values of the position and momentum variances. The purity then quickly decays in the time interval $\Delta t $ [in a timescale of order $(\gamma_m n_b)^{-1}$] due to the thermalization process. We have also verified that this behavior is well reproduced by the solutions of the coupled set of deterministic equations, Eqs.~(\ref{eq:semiclassicaltris}) and Eq.~(\ref{compact2}) of Sec.~III.

After verifying that the properly calibrated probe output phase quadrature $Y_2^{\rm out}(t)$ correctly reproduces the mechanical dynamics, we have taken the data in the time interval from $t_1$ to $t_2$, and performed a fast Fourier transform (FFT) of $Y_2^{\rm out}(t)$, obtaining the output spectrum $S_{Y_2Y_2}^{\rm out}(\omega)$ shown in Fig.~\ref{fig1March}(c). The relevant quantities in this spectrum are: i) the heights of the calibration and of the resonant signal peak, from which we get the calibrated measured value for $|A_b|^2$ in the FFT time interval $\Delta t$; ii) the central frequency of the signal peak $\omega_m'$, from which we get the normalized frequency shift $(\omega_m'-\omega_m)/\omega_m$. In fact, as Eq.~(\ref{abtildesolinst}) suggests, the ratio $(\omega_m'-\omega_m)/\omega_m |A_b|^2$ provides an estimate of $\beta_{NL}$.

Fig.~\ref{fig1March}(b) shows that the mechanical state purity quickly decays within the FFT integration time $\Delta t$ due to the thermalization process. Therefore the estimation of $\beta_{NL}$ is influenced not only by quantum fluctuations but also by thermal ones. However, we have verified that we cannot take a shorter FFT time interval without compromising the frequency resolution needed to resolve the nonlinear frequency shift. We can say therefore that this is the best quantum estimation of $\beta_{NL}$ we can make for the chosen set of parameters. 

\begin{figure}[h]
	\centering
\includegraphics[width=.99\linewidth]{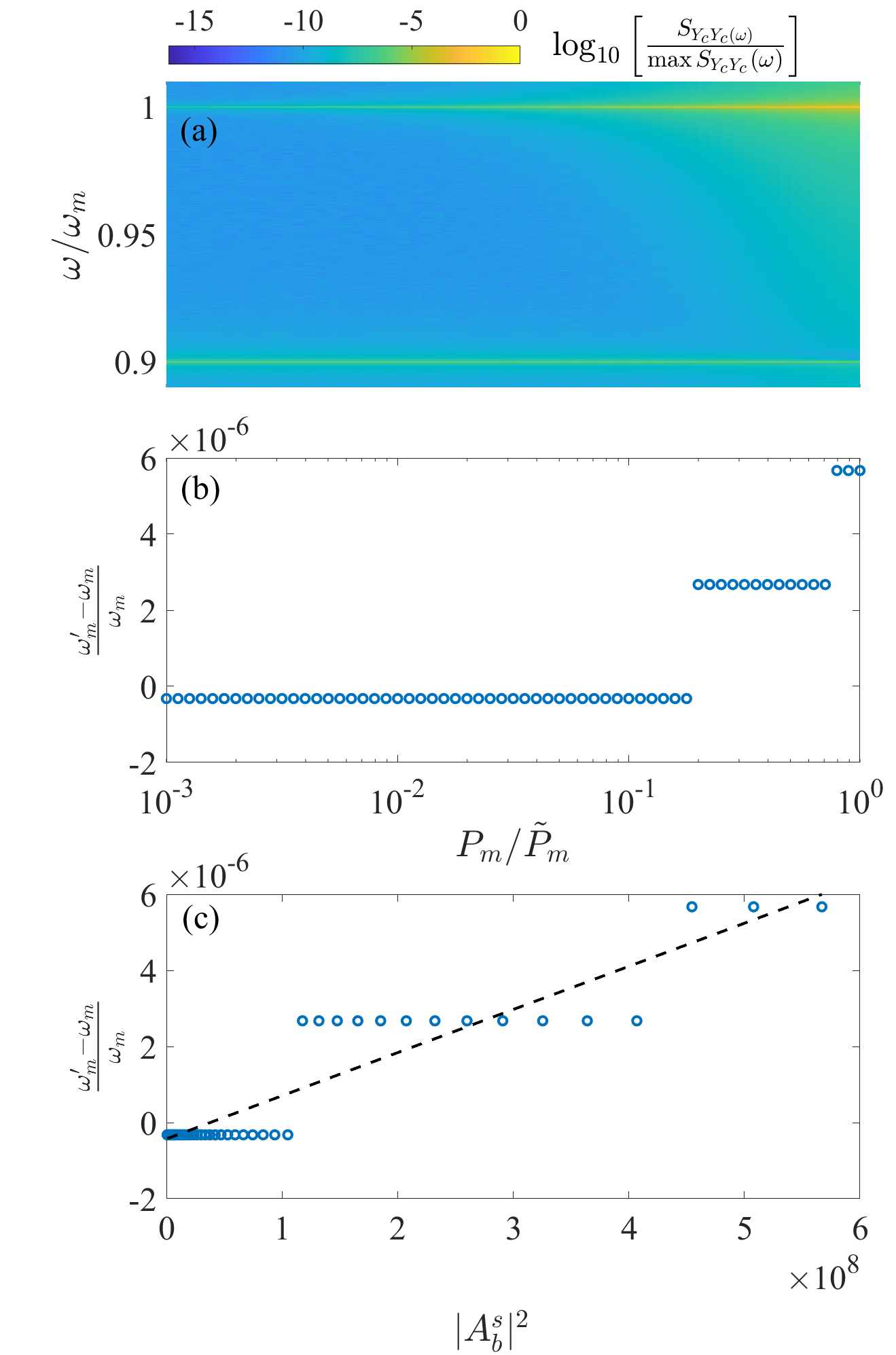}
\caption{(a) Output probe homodyne spectra $S_{Y_2Y_2}(\omega)$ versus the normalized modulation pump power $P_m/\tilde{P}_m$ (where $\tilde{P}_m = 2 \times 10^{-6}$ W). (b) Normalized nonlinear frequency shift versus $P_m/\tilde{P}_m$. (c) Normalized frequency shift versus the corresponding calibrated value $|A_b|^2$; the dashed line is the linear fit whose slope provides the estimated value of $\beta_{NL}$ (which is $\beta_{NL}^{est}=1.13 \times 10^{-14}$ in this case). The parameters are the same of Fig. \protect\ref{fig:spectrapeak}.
	}
\label{fig3March}
\end{figure}

In order to have a robust estimation for $\beta_{NL}$ we then repeat the same numerical analysis of Fig.~\ref{fig1March} for different values of the modulation power $P_m$, corresponding to different values of $E_m$ and of the amplitude of the prepared coherent state $|A_b^s|$. The results are collectively shown in Fig.~\ref{fig3March}. In Fig.~\ref{fig3March}(a) we plot the probe output spectra as a function of the modulation power $P_m$, while the corresponding normalized frequency shift $(\omega_m'-\omega_m)/\omega_m $ versus $P_m$ is shown in Fig.~\ref{fig3March}(b). Then Fig.~\ref{fig3March}(c) shows the normalized frequency shift versus the corresponding calibrated value $|A_b|^2$, while the dashed line is the linear fit whose slope provides the estimated value of $\beta_{NL}$. These plots show a step behavior of the normalized frequency shift because its values, due to the very small value of the nonlinearity, are very close to the frequency resolution given by the numerically evaluated FFT.

\begin{figure}[h]
	\centering	\includegraphics[width=.99\linewidth]{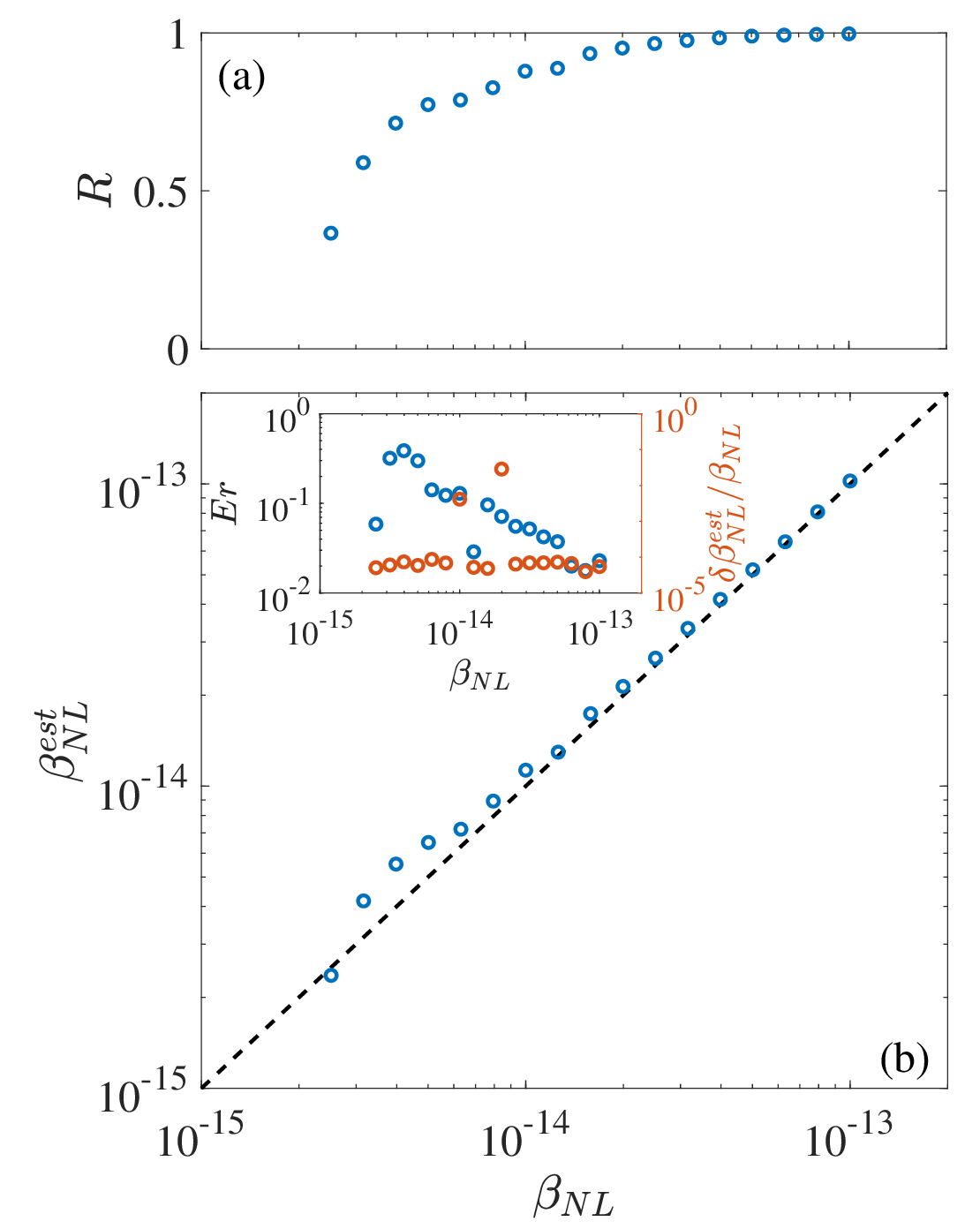}
	\caption{
In (a) the blue dots represent the linear regression coefficient $R$ of the fitting process for each value of $\beta_{NL}$.
(b) Plot of the estimated $\beta_{NL}^{est}$ versus $\beta_{NL}$. Each point corresponds to the average value of 13 simulations. In the inset of (b) the blue dots correspond to the relative error $E_r=|\beta_{NL}^{est}-\beta_{NL}|/\beta_{NL}$ (left axis),  while the red dots represent the relative fluctuation $\delta\beta_{NL}^{est}/\beta_{NL}$ of these 13 simulations (right axis). 
}
\label{fig4March}
\end{figure}

Figs.~\ref{fig1March} and \ref{fig3March} describe the protocol for estimating the weak nonlinear parameter $\beta_{NL}$. In order to establish the sensitivity and robustness of this protocol, we have repeated the same numerical analysis for many different values of $\beta_{NL}$. The corresponding results, for $\beta_{NL} \in [2.5 \times 10^{-15}, 10^{-13}]$, are summarized in Fig.~\ref{fig4March}. 
In (a) the blue dots represent the linear regression coefficient $R$
of the fitting process, for each value of $\beta_{NL}$.
In (b) we plot the estimated $\beta_{NL}^{est}$ versus $\beta_{NL}$. We see that the estimation protocol is reliable and consistent in the chosen parameter region; moreover, we see that its sensitivity, that is, the smallest nonlinear coefficient that our protocol is able to estimate, is $\beta_{NL}^{min} \simeq 2.5 \times 10^{-15}$, which is related to the precision of the present numerical analysis. Each point here corresponds to the average value of 13 simulations. In the inset of (b) the blue dots correspond to the relative error $E_r=|\beta_{NL}^{est}-\beta_{NL}|/\beta_{NL}$ (left axis),  while the red dots are the relative fluctuation $\delta\beta_{NL}^{est}/\beta_{NL}$ of these 13 simulations. For smaller values of $\beta_{NL}$ the nonlinear frequency shift is smaller, and the estimation error $E_r$ becomes appreciably larger than the statistical error, because the finite frequency resolution of the numerical simulation tends to yield an increasing systematic error.

Our numerical analysis shows that the nonstationary scheme presented here is reliable and significantly more sensitive than the stationary scheme of Sec. IV. Our analysis also suggests that the sensitivity achievable for $\beta_{NL}$ in an experiment limited only by quantum zero-point fluctuations, with a sideband-cooled resonator in a cavity, is unable to reach the sensitivity $\beta_{NL} \simeq 4 \times 10^{-21}$ achieved with similar membranes in a classical scenario in Ref.~\cite{Bawaj2015}. In fact, this experiment, as well as those of Refs.~\cite{bushev,tobar}, fully exploits the amplification provided by large values of $|A_b(t)|$, which are possible with an interferometric readout in a fully classical regime in the presence of large thermal noise. Instead, operating with large amplitude mechanical oscillations within a cavity becomes nontrivial as soon as $\xi_j \simeq g_j |A_b|/\omega_m > 1$, because the mechanically-induced cavity frequency modulation becomes large compared to the cavity linewidth (especially in the resolved sideband regime), and it becomes increasingly harder to keep the cavity locked (see e.g., Fig. 3 and the corresponding description in Ref.~\cite{Piergentili2021}). Despite the lower sensitivity, it is still of fundamental importance to test potential gravity effects such as those associated with deformed commutators \cite{Pikovski,Bawaj2015,bushev,bonaldi,kumar,tobar,bose,marletto}, or nonlocal approaches to quantum gravity \cite{belenchia,Wang} in a quantum regime. In fact, only in this regime one can probe the effect of quantum fluctuations and quantum indeterminacy on the gravitational field.

\section{Concluding remarks}

We have described in detail various strategies for the sensitive measurement of weak nonlinearities of a mechanical resonator in a regime dominated by quantum fluctuations. These schemes are particularly relevant for probing new physics which is responsible for the appearance of weak effective mechanical nonlinearities, such as those associated with the nonrelativistic limit of some quantum gravity theories \cite{Pikovski,Bawaj2015,bushev,bonaldi,kumar,tobar,bose,marletto} or nonlocal approaches to quantum gravity \cite{belenchia,Wang}. 

We propose large-amplitude pure coherent states of a mechanical resonator as a powerful tool for providing such an estimation of the effective nonlinear parameter. We first consider the generation of this state in a stationary regime through a driven version of the standard sideband cooling protocol \cite{Aspelmeyer2014,Genes2009}. However, this stationary estimation of the nonlinearity is not very sensitive because it tends to be hidden by the simultaneous and unavoidable presence of the radiation pressure effects associated with the ground state cooling process. 

We then consider a nonstationary strategy in which the generated large-amplitude almost pure coherent state slowly decays and thermalizes to the equilibrium thermal state, because the cooling pump and its modulation are turned off. 
We have shown that, if we monitor this mechanical decay with a probe mode exactly at resonance with its cavity mode, and we employ a high-quality factor resonator in a cryogenic environment, one can get a sensitive estimation of the nonlinearity parameter $\beta_{NL}$ in a regime influenced by quantum fluctuations only. This nonstationary method represents the quantum version, within an optomechanical cavity, of the method applied in Refs. \cite{Bawaj2015,bushev,tobar} using macroscopic resonators with a fully classical dynamics. The sensitivity achievable in this quantum version is however worser, due to the difficulty of operating with very large mechanical amplitudes within a cavity in the resolved sideband regime, which are instead possible in the classical regime. Nonetheless, only an experiment carried out within a quantum regime is able to give some information on the eventual effects of quantum fluctuations and quantum indeterminacy on gravity.

For completeness we point out other additional elements that one has to take into account for a successful implementation of the nonlinearity estimation protocol (see also Ref.~\cite{bonaldi}).

First, the fast transient soon after turning off the cooling pump and its modulation certainly disturbs the PDH locking system, which needs some time to work properly again. 

Then, there are various limitations and technical difficulties associated with the presence of nearby mechanical modes of our resonator. (i) The sudden variation of the static radiation pressure force due to the pump switch-off, responsible for the change $\beta_0 \to \beta_0'$, acts on \emph{all} mechanical modes, and it is larger for the low frequency mechanical modes (see Eq.~(\ref{approxbeta0})). As a consequence, \emph{all} the other mechanical modes are also excited and this may disturb the observation of the dynamics of the resonator of interest. In this respect it is convenient to perform the experiment with the lower frequency, fundamental vibrational mode. 
(ii) The heating associated with the thermalization process affects \emph{all} mechanical modes, and the nearby modes may increase significantly the background noise.

\begin{acknowledgements}
We acknowledge financial support from NQSTI within PNRR MUR Project PE0000023-NQSTI.
\end{acknowledgements}

\appendix

\section{Alternative nonstationary strategy: turning off the pump modulation only}\label{intermediate}

We can also consider an alternative nonstationary strategy, which is intermediate between the two discussed in the main text. In fact, after reaching the steady state of Sec.~\ref{stationary}, one may turn off only the modulation of the pump, which implies taking $E_m=0$ in Eq.~(\ref{eq:quantum langevin1}) and Eq.~(\ref{eq:semiclassicalbis 1}). The corresponding dynamics is again nonstationary, but different from the case when also the pump driving is turned off. Here we assume for simplicity that the modulation is turned-off instantaneously.

The dynamics of the amplitudes in this latter case can be solved by adapting the results of Sec.~\ref{pumpoff} to the limiting case $\tau \to \infty$, i.e., assuming that the pump drive does not decay anymore. One can closely follow the same steps, with few, but relevant differences: i) the constant shift $\beta_0$ does not change, and therefore there is no effect on the detunings and no adjustment needed from the PDH+servo loop locking system. ii) There is again a transient and a stationary term in the cavity amplitudes $\alpha_j$ but they are different from those of Sec.~\ref{pumpoff}. In fact, Eq.~(\ref{eq:cavity field solution_tranrewri}) becomes
\begin{eqnarray}
&&\alpha_j(t)  =  e^{i\psi_j(t)}\left\{\alpha_j^{\rm trans}(t)\right. \label{eq:cavity field solution_tranrewri2}  \\
&& \left.+ E_j \sum_{n=-\infty}^{\infty}\dfrac{J_n\left(-\tilde{\xi}_j\right)e^{i(\omega_m t-\theta)n}}{i\omega_m n-\mathcal{L}_j} \right\},  \nonumber
\end{eqnarray}
where the initial transient term now decays with the cavity decay times $\kappa_j$ and it is given by
\begin{eqnarray}
&&\alpha_j^{\rm trans}(t)  =  e^{\mathcal{L}_j t}\left\{\right. \alpha_j(0)e^{i\phi_j^{\rm fast}-i\psi_j(0)} \label{eq:cavity field solution_tran_tran2}\\
&&- \left. E_j \sum_{n=-\infty}^{\infty}\dfrac{J_n\left(-\tilde{\xi}_j\right)e^{-i\theta n}}{i\omega_m n-\mathcal{L}_j}\right\} .
\end{eqnarray}
One can then follow the same steps of Sec.~\ref{pumpoff} and arrive at the relevant equation for the slowly varying amplitude $A_b(t)$, which again will decay from the initial state generated by the modulation, and will be also influenced by the mechanical and radiation pressure nonlinearities. The final equation is a modified version of
Eq.~(\ref{eq:first-order solution_tran2}), including also the effect of the pump,
\begin{eqnarray}
&&\dot{A}_b(t)=\left(-\gamma_m-i\Delta_m'\right) A_b(t)-i\beta_{\rm NL}\omega_m |A_b(t)|^2 A_b(t) \nonumber \\
&&+i \sum_{j=1,2}\sum_{n=-\infty}^{\infty}\dfrac{g_j E_j^2 e^{i\theta} J_n\left(-\tilde{\xi}_j\right)J_{n+1}\left(-\tilde{\xi}_j\right)}{(i\omega_m n-\mathcal{L}_j)[-i\omega_m (n+1)-\mathcal{L}_j^*]} .
\label{eq:first-order solution_tran2bis}
\end{eqnarray}
However, now the radiation pressure terms described by the sum over the Bessel functions, cannot be completely eliminated as in the case without the pump. In fact, one can take again a perfectly resonant probe driving $\Delta_2 =0$ and therefore eliminate at all orders the optomechanical effect of the probe mode, but this does not occur for the pump mode ($j=1$) which is red detuned and is responsible for the cooling of the mechanical resonator.

Therefore we now have a situation similar to that of Sec. \ref{stationary}A, and we can arrive at a similar amplitude equation for $A_b(t)$ (even though now referred to a carrier frequency at $\omega_m$ rather than $\delta_m$). In order to describe the effect of the radiation pressure nonlinearity we develop the sum terms in Eq.~(\ref{eq:first-order solution_tran2bis}) in series of powers of $\tilde{\xi}_j$ and stop at third order. We finally get an evolution equation similar to Eq.~(\ref{eq:first-order solution_appr}),
\begin{equation}
\dot{A}_b(t)=\left(-\gamma_m^{\rm eff}-i\Delta_m^{\rm eff}\right) A_b(t)+\left(c_3-i\beta_{\rm NL}\omega_m \right) |A_b(t)|^2 A_b(t),
\label{eq:first-order solution_appr2}
\end{equation}
with the effective damping and frequency shift parameters
\begin{eqnarray}
  \gamma_m^{\rm eff} &=& \gamma_m-{\rm Re}(c_1), \\
  \Delta_m^{\rm eff} &=& \Delta_m'-{\rm Im}(c_1),
\end{eqnarray}
and with
\begin{eqnarray}
  c_1  &=& i\sum_{j=1,2}\frac{g_j^2  E_j^2}{\omega_m}\left[\frac{1}{(\kappa_j+i\Delta_j)[\kappa_j-i(\omega_m+\Delta_j)]}\right. \nonumber \\
  &&\left.-\frac{1}{(\kappa_j-i\Delta_j)[\kappa_j-i(\omega_m-\Delta_j)]}\right], \\
   c_3 \hspace{-3pt} &=& \hspace{-3pt} i\sum_{j=1,2}\frac{g_j^4  E_j^2}{2\omega_m^3}\left[\frac{1}{[\kappa_j+i(\omega_m+\Delta_j)][\kappa_j-i(2\omega_m+\Delta_j)]}\right. \nonumber \\
  &&\left.-\frac{1}{[\kappa_j+i(\omega_m-\Delta_j)][\kappa_j-i(2\omega_m-\Delta_j)]}\right].
\end{eqnarray}
In the expression for the coefficients $c_1$ and $c_3$ we have kept the probe terms (the $j=2$ terms in the sums) for generality but, as we have seen in the previous Section, these terms are zero at all orders if we can take $\Delta_2=0$ exactly.

Eq.~(\ref{eq:first-order solution_appr2}) is of the same form of Eq.~(\ref{eq:first-order solution_tran2}) and therefore it can be exactly solved using steps similar to those used in Sec.~\ref{pumpoff}B, the main difference being that $c_3$ has in general a nonzero real part, implying the presence of an additional nonlinear damping term.

We rewrite again
\begin{equation}\label{abtildeinter}
  A_b(t)=\tilde{A}_b(t)e^{-(\gamma_m^{\rm eff}+i \Delta_m^{\rm eff})t},
\end{equation}
so that Eq.~(\ref{eq:first-order solution_appr2}) yields the simpler equation for $\tilde{A}_b(t)$
\begin{equation}\label{abtildeeq}
 \dot{\tilde{A}}_b(t)=\left(c_3-i\beta_{\rm NL}\omega_m\right) e^{-2\gamma_m^{\rm eff} t}\vert\tilde{A}_b(t)\vert^2 \tilde{A}_b(t).
\end{equation}
Rewriting it as an equation for modulus and phase, $ \tilde{A}_b(t)=\vert\tilde{A}_b(t)\vert e^{i\tilde{\theta}(t)}=r(t)e^{i\tilde{\theta}(t)}$, we see that the modulus $r$ is not constant anymore, but it satisfies the equation
\begin{equation}\label{eqdiffmodu}
  \dot{r}(t)={\rm Re}(c_3) e^{-2\gamma_m^{\rm eff} t} r^3(t),
\end{equation}
with solution
\begin{equation}\label{eqfiffmodusol}
  r^2(t)=\frac{r(0)^2}{1-r(0)^2\left[{\rm Re}(c_3)/\gamma_m^{\rm eff}\right]\left(1-e^{-2\gamma_m^{\rm eff} t}\right)},
\end{equation}
which has to be inserted into the equation for the phase $\tilde{\theta}(t)$
\begin{equation}\label{thetatilde2}
\dot{\tilde{\theta}}(t)= \left[{\rm Im}(c_3)-\beta_{\rm NL}\omega_m \right]e^{-2\gamma_m^{\rm eff} t}r^2(t),
\end{equation}
giving the solution
\begin{eqnarray}\label{thetatildesol}
 && \tilde{\theta}(t)= \theta_0 -\frac{\left[{\rm Im}(c_3)-\beta_{\rm NL}\omega_m\right]}{2{\rm Re}(c_3)} \\
 && \times \ln\left[1-\vert A_b(0)\vert^2 \frac{{\rm Re}(c_3)}{\gamma_m^{\rm eff}}\left(1-e^{-2\gamma_m^{\rm eff} t}\right)\right]. \nonumber
\end{eqnarray}
We finally get
\begin{equation}\label{abtildesol2}
A_b(t)=\frac{A_b(0)e^{-(\gamma_m^{\rm eff}+i \Delta_m^{\rm eff})t}}{\left[1-\vert A_b(0)\vert^2 \frac{{\rm Re}(c_3)}{\gamma_m^{\rm eff}}\left(1-e^{-2\gamma_m^{\rm eff} t}\right)\right]^{\frac{1}{2}+i\frac{\left[{\rm Im}(c_3)-\beta_{\rm NL}\omega_m\right]}{2{\rm Re}(c_3)}}},
\end{equation}
which means that the mechanical oscillator decays with rate $\gamma_m^{\rm eff}$, which is faster than the rate of the case without the pump drive by a factor roughly corresponding to the optomechanical cooperativity, and with an amplitude-dependent instantaneous frequency
\begin{eqnarray}\label{freqinst2}
 && \omega_{\rm inst}= \omega_m+\Delta_m^{\rm eff}-\dot{\tilde{\theta}}\\
 &&=\omega_m+\Delta_m^{\rm eff}+\frac{\left(\beta_{\rm NL}\omega_m -{\rm Im}(c_3)\right)e^{-2\gamma_m^{\rm eff} t}\vert A_b(0)\vert^2}{1-\vert A_b(0)\vert^2\left[{\rm Re}(c_3)/\gamma_m^{\rm eff}\right]\left(1-e^{-2\gamma_m^{\rm eff} t}\right)}. \nonumber
\end{eqnarray}
Also in this last case one can get an estimation of $\beta_{\rm NL}$, by looking at the dependence of instantaneous frequency versus the initial squared amplitude $\vert A_b(0)\vert^2$. 

\begin{figure}[h!]
	\centering
\includegraphics[width=.99\linewidth]{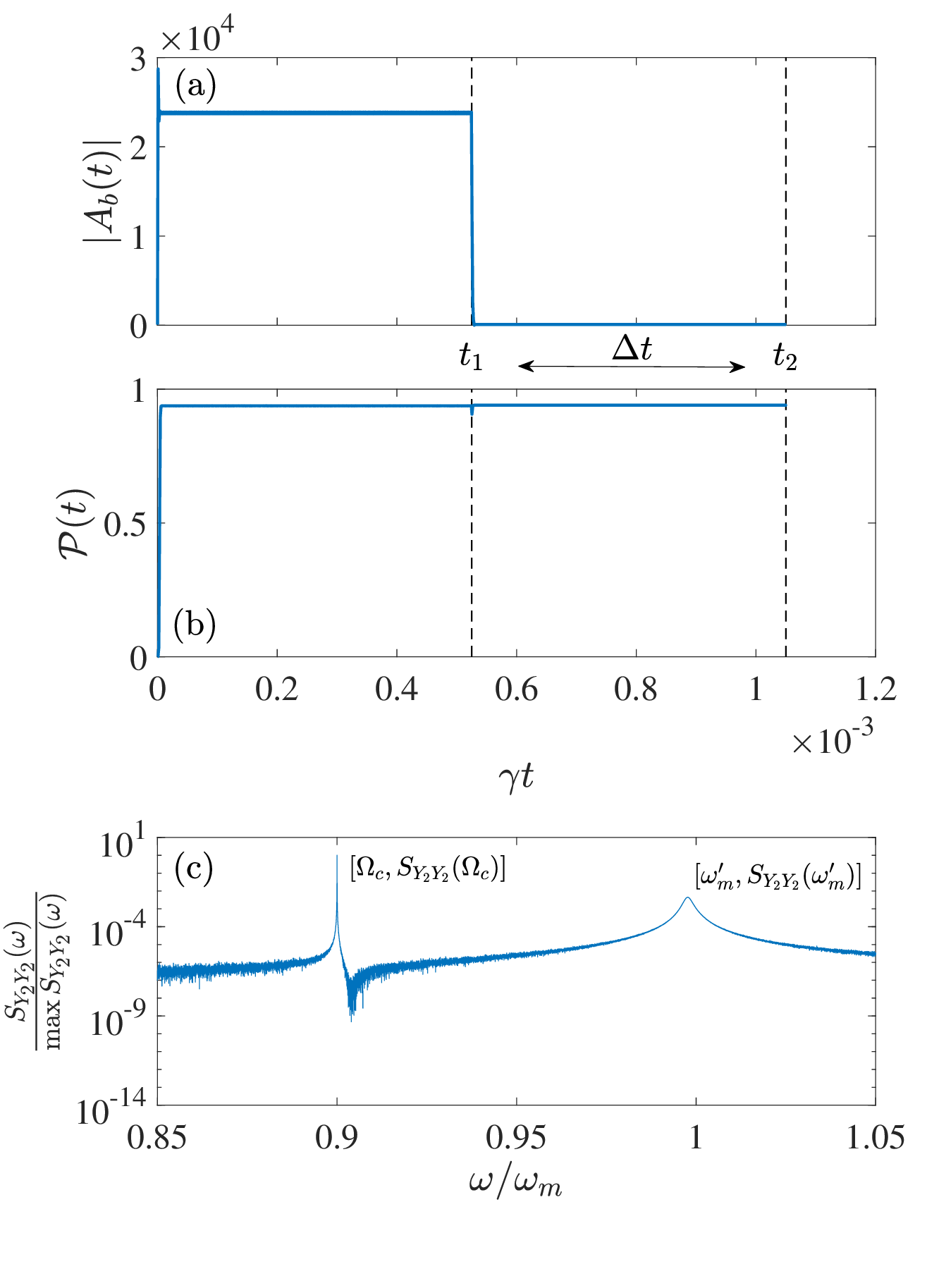}
\caption{Time evolution of the coherent amplitude $|A_b(t)|$ (a), and of the purity ${\cal P}(t)$ (b), of the mechanical reduced state. In the time interval from $t=0$ to $t=t_1$ a very pure large-amplitude coherent state is prepared. At $t=t_1$ only the modulation tone is turned off: $|A_b(t)|$ quickly decays to zero, while ${\cal P}(t)$ remains unchanged. In (c) we plot the output probe homodyne spectrum $S_{Y_2Y_2}(\omega)$ obtained via a FFT over the time interval $\Delta t = t_2-t_1 = 5.25 \times 10^{-4}\gamma_m^{-1}$ The calibration peak and the resonant signal peak are again visible, but now the resonant peak is much wider and lower. Parameters are the same as those of Fig.~\protect\ref{fig1March}.
	}
\label{fig5March}
\end{figure}

\begin{figure}[h!]
	\centering
\includegraphics[width=.99\linewidth]{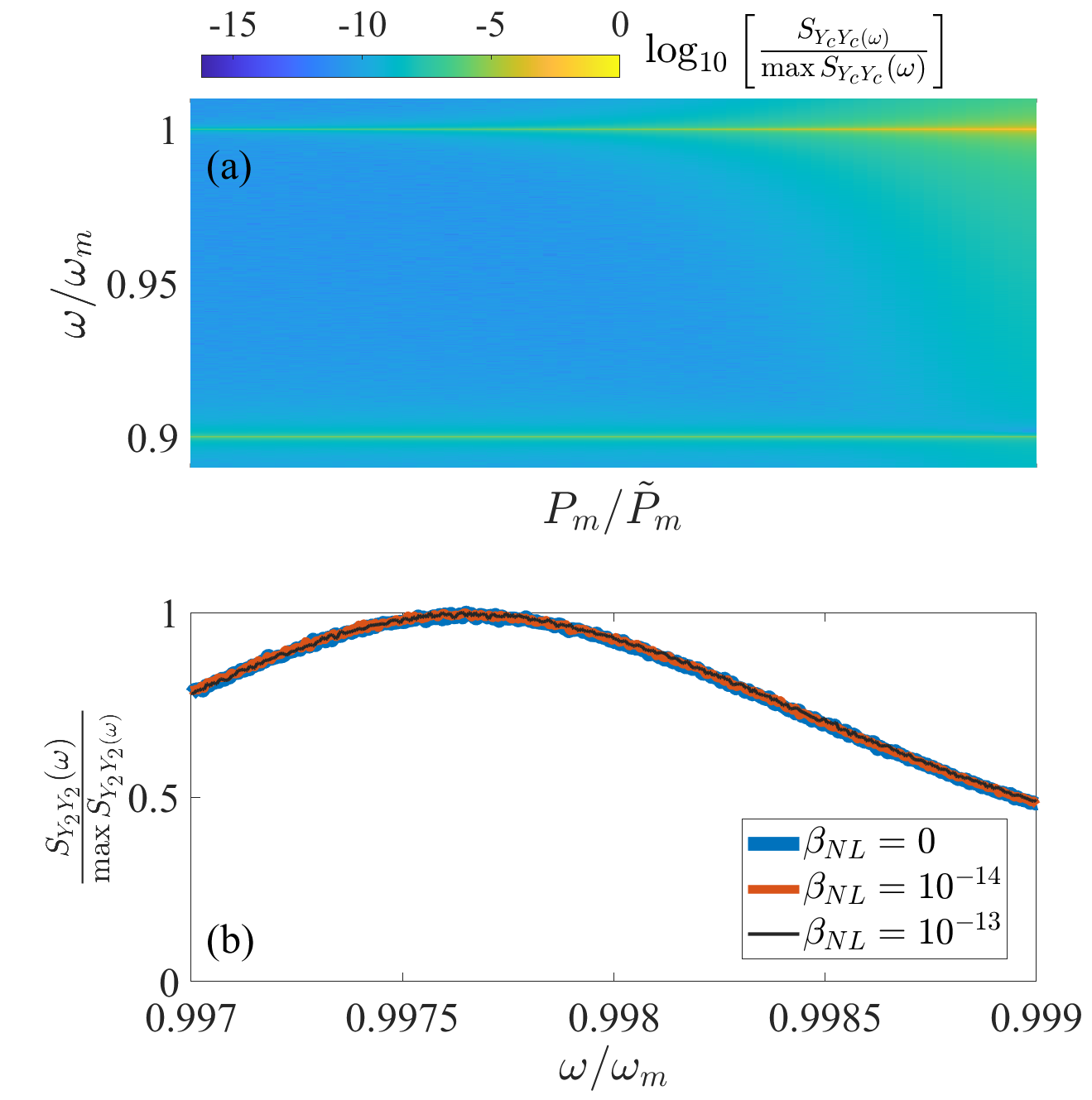}
\caption{(a) Output probe homodyne spectra $S_{Y_2Y_2}(\omega)$ versus the normalized power of the modulation at $\delta_m$, $P_m/\tilde{P}_m$ (where $\tilde{P}_m=2 \times 10^{-6}$ W). (b) Comparison between the output spectra at different values of the nonlinear parameter $\beta_{NL}$, and with fixed value of the modulation power, $P_m = \tilde{P}_m$. The plot shows that the protocol has a sensitivity $\beta_{NL}^{min}$ worser than $10^{-13}$. The other parameters are the same of Fig.~\protect\ref{fig5March}.
	}
\label{fig6March}
\end{figure}

We can again look at the time evolution obtained by averaging the trajectories obtained from the simulated Langevin equations, and repeat the analysis of Fig.~\ref{fig1March} for the same set of parameters. This is shown in Fig.~\ref{fig5March}: Fig.~\ref{fig5March}(a) shows that the amplitude $|A_b(t)|$ now quickly decays to zero due to the large effective damping $\gamma_m^{\rm eff}$ of the ground-state-cooled mechanical resonator. On the contrary, in Fig.~\ref{fig5March}(b) we see that the purity ${\cal P}(t)$ remains stable and very close to one,  thanks to the cooling laser, i.e., the regime is always dominated by the quantum zero-point fluctuations. 
The homodyne probe output spectrum  $S_{Y_2Y_2}^{\rm out}(\omega)$ obtained from the FFT of $Y_2^{\rm out}(t)$, in the same time interval from $t_1$ to $t_2$ is then shown in Fig.~\ref{fig5March}(c). 
This last plot however shows why the estimation of nonlinearity is seriously hindered in this nonstationary strategy. In fact, $A_b(t)$ now decays with the much faster rate $\gamma_m^{\rm eff}$, and the resonant peak in the corresponding output spectrum is now much lower and wider. This makes any estimation of the nonlinear frequency shift almost impossible. This is shown in Fig.~\ref{fig6March}, where we plot the output homodyne spectrum versus the power of the modulation, Fig.~\ref{fig6March}(a), and, at fixed power, for three different values of the nonlinearity, Fig.~\ref{fig6March}(b). We see that the protocol is unable to discriminate between the case with $\beta_{NL}=0$ and $\beta_{NL}=10^{-13}$. The impossibility to resolve the frequency shift also hides the eventual bias in the estimation due to the presence of the nonzero radiation pressure nonlinearity coefficient $c_3$.  Therefore, keeping the cooling drive on allows staying within the fully quantum regime, but the very large effective damping makes the sensitive estimation of $\beta_{NL}$ impossible.

\end{document}